\documentclass[12pt]{article}

\usepackage{amsmath,graphicx}
\usepackage{multirow}
\usepackage{amsfonts}
\usepackage{amssymb}
\usepackage{amscd}
\usepackage{cite}
\usepackage{amsmath}
\usepackage{bbm}

\def\hybrid{\topmargin -20pt    \oddsidemargin 0pt
        \headheight 0pt \headsep 0pt
        \textwidth 6.25in       
        \textheight 9.5in       
        \marginparwidth .875in
        \parskip 5pt plus 1pt   \jot = 1.5ex}

\hybrid

\numberwithin{equation}{section}
\numberwithin{table}{section}

\newcommand{\be}{\begin{equation}}
\newcommand{\ee}{\end{equation}}
\newcommand{\M}{{\bf M}}
\newcommand{\bea}{\begin{eqnarray}}
\newcommand{\eea}{\end{eqnarray}}

\def\cN{{\cal N}}
\def\cF{{\cal F}}

\def\nv{n_{\rm v}}
\def\nh{n_{\rm h}}
\renewcommand{\Re}{\operatorname{Re}}
\renewcommand{\Im}{\operatorname{Im}}
\newcommand\e{\mathrm{e}}
\newcommand\iu{\operatorname{i}}
\newcommand\diff{\mathrm{d}}

\newcommand{\pt}{\partial}

\def\ax{{\tilde \phi}} 

\def\mino{m_{3/2}}
\def\k{\hat{k}}
\def\h{\hat{h}}
\def\K{\hat{K}}
\def\Mq{\hat{\M}}

\def\J{\hat{J}}
\def\kk{{k}}
\def\mm{{\tilde m}}
\def\Proj{{\Pi}}

\def\C{C}    

\def\PQ{\tilde \pi}

\begin{document}
\begin{titlepage}
\begin{center}
\rightline{\small ZMP-HH/10-5}
\vskip 1cm

{\Large \bf
The $\cN=1$ Low-Energy Effective Action of Spontaneously Broken  $\cN=2$
Supergravities}
\vskip 1.2cm
{\bf Jan Louis$^{a,b}$, Paul Smyth$^{a}$ and Hagen Triendl$^{a}$}

\vskip 0.8cm

$^{a}${\em II. Institut f\"ur Theoretische Physik der Universit\"at Hamburg, Luruper Chaussee 149, 22761 Hamburg, Germany}
\vskip 0.4cm

{}$^{b}${\em Zentrum f\"ur Mathematische Physik,
Universit\"at Hamburg,\\
Bundesstrasse 55, D-20146 Hamburg}
\vskip 0.8cm

{\tt jan.louis,paul.smyth,hagen.triendl@desy.de}

\end{center}

\vskip 1cm

\begin{center} {\bf ABSTRACT } \end{center}

\noindent

We consider the class of four-dimensional $\cN=2$ gauged supergravities whose maximally symmetric ground states leave only one of the two supersymmetries intact.
For these theories we derive
the low-energy effective action below the scale of partial
supersymmetry breaking and compute the  $\cN = 1 $ couplings in terms
of the $\cN=2$ `input data'. We show that this effective action
satisfies the constraints of $\cN = 1 $ supergravity in that its
$\sigma$-model metric is K\"ahler, while the superpotential and the gauge
kinetic functions are holomorphic.
As an example we discuss the $\cN=1$ effective supergravity of type II compactifications.

\bigskip

\vfill
August, 2010


\end{titlepage}

\section{Introduction}
In a recent paper \cite{Louis:2009xd} we discussed spontaneous $\cN=2
\to \cN=1$ supersymmetry breaking in four-dimensional supergravity and type II string compactifications using the embedding tensor formalism
\cite{deWit:2002vt,deWit:2005ub}. We confirmed that the simultaneous
appearance of electric and magnetic charges is necessary to circumvent
the old no-go theorem forbidding partial $\cN=2 \to \cN=1$
supersymmetry breaking in theories with only electric charges \cite{Cecotti:1984rk,Cecotti:1984wn}, analogous to the case of rigid
supersymmetry \cite{Antoniadis:1995vb}. This fact is particularly
transparent in the embedding tensor formalism which treats electric
gauge bosons and their magnetic duals on the same footing.

Specific examples of supergravity theories which display partial
supersymmetry breaking have been presented in \cite{Cecotti:1985sf,Ferrara:1995gu,Ferrara:1995xi,Fre:1996js}, generalising
the mechanism of adding a magnetic Fayet-Illiopoulos term to a rigid
supersymmetric theory \cite{Antoniadis:1995vb}.\footnote{For an analogous discussion in string theory see, for example, \cite{Kiritsis:1997ca}.} In \cite{Louis:2009xd}
we adopted a more general approach, in that we analysed arbitrary
$\cN=2$ gauged supergravities and showed that the conditions for
partial supersymmetry breaking in a maximally symmetric background primarily determine the structure of
the embedding tensor, i.e.\ the spectrum of electric and magnetic
charges, but do not constrain the
scalar field space ${\M}_{\rm v}$ of the vector multiplets.
In the hypermultiplet sector on the other hand,
the scalar field space ${\M}_{\rm h}$ has to admit
at least two linearly independent, commuting
isometries. It is necessary to gauge these isometries in order to induce masses for the two Abelian gauge
bosons which join the heavy gravitino in a massive $\cN=1$ gravitino
multiplet.  Partial supersymmetry breaking further demands that a
specific linear combination of the two Killing vectors generating the isometries
is holomorphic with respect to one of the three almost complex
structures which exist on ${\M}_{\rm h}$.  In \cite{Louis:2009xd} we
explicitly identified two such Killing vectors for the specific class
of special quaternionic-K\"ahler manifolds \cite{Cecotti:1988qn}. These
manifolds are in the image of the c-map and so arise at tree-level in type II compactifications on Calabi-Yau or generalised manifolds
with $SU(3)\times SU(3)$ structure
\cite{Cecotti:1988qn,Gurrieri:2002wz,D'Auria:2004tr,Grana:2005ny,Grana:2006hr,Cassani:2007pq,Cassani:2008rb,Grana:2009im}.
However, in this paper we shall keep the discussion more general and
discuss partial supersymmetry breaking in generic $\cN=2$ supergravities.
The special quaternionic-K\"ahler manifolds then
serve as a convenient explicit example.

The aim of the present paper is to continue the analysis of \cite{Louis:2009xd} and
derive the $\cN=1$ low-energy effective
action that is valid below the scale of partial supersymmetry breaking $\mino$
or, in other words, below the scale set by the heavy gravitino.
In order to achieve this we integrate out the entire massive
$\cN=1$ gravitino multiplet (containing fields with spin $s=(3/2,1,1,1/2)$)
together with
all other multiplets which, due to the symmetry breaking, acquire masses
of ${\cal O}(\mino)$.  This results in
an effective $\cN=1$ theory whose couplings are determined by the
couplings of the `parent' $\cN=2$ theory.\footnote{Preliminary aspects of this programme were presented in \cite{Louis:2002vy,Gunara:2003td}.}

An interesting aspect of the effective theory is the structure of the
scalar field space~$\M$. In $\cN=2$ supergravities~$\M$ is a direct
product of the form  \cite{Bagger:1983tt,deWit:1984pk,deWit:1984px,Andrianopoli:1996cm,Craps:1997gp}
\begin{equation}
  \label{N=2product}
  \M \ =\ {\M}_{\rm  h} \times {\M}_{\rm  v}\ ,
\end{equation}
where ${\M}_{\rm h}$ is the $4 n_{\rm h}$-dimensional
quaternionic-K\"ahler manifold spanned by the scalars of $n_{\rm h}$
hypermultiplets, while ${\M}_{\rm v}$ is a $2 n_{\rm v}$-dimensional
special-K\"ahler manifold spanned by the scalars of $n_{\rm v}$ vector
multiplets. Note that ${\M}_{\rm v}$ is a K\"ahler manifold but ${\M}_{\rm h}$
is not.  We shall see that the process of integrating out the two heavy
gauge bosons corresponds to taking the quotient of
${\M}_{\rm h}$ with respect to the two isometries generating the partial supersymmetry breaking.  This leaves a $(4
n_{\rm h}-2)$-dimensional manifold $\Mq_{\rm h}$ where the two
`missing' scalar fields are the Goldstone bosons eaten by the heavy
gauge bosons. We shall show that $\Mq_{\rm h}$ is equipped with a
K\"ahler metric consistent with the $\cN=1$ supersymmetry of the low-energy effective theory.\footnote{A more detailed analysis of the mathematical properties of this construction will be presented in a companion paper \cite{Cortes}.} It is also possible that, apart from the two gauge bosons, other scalar fields (from both vector- and
hypermultiplets) acquire a mass of ${\cal O}(m_{3/2})$ and thus
have to be integrated out, leading to a further reduction of the scalar field space. However, as such scalars are not Goldstone bosons this process simply amounts to projecting to a K\"ahler submanifold of
$\Mq_{\rm h}\times {\M}_{\rm v}$, rather than taking a quotient.  The resulting $\cN=1$ scalar field
space is then given by
\begin{equation}
  \label{N=1product}
 \M^{\cN=1} = {\Mq}_{\rm  h} \times {\Mq}_{\rm  v}\ ,
 \end{equation}
where ${\Mq}_{\rm v}$ is a submanifold of ${\M}_{\rm v}$. (For
notational simplicity we did not introduce a new symbol for the
submanifold of $\Mq_{\rm h}$.)

The dimension of  $\M^{\cN=1}$ is model dependent.
It can be as large as $2n_{\rm v} + 4 n_{\rm h}-2$
when the only scalars integrated out are the two Goldstone bosons providing the mass degrees of freedom for the heavy gauge bosons.
However, the dimension of  $\M^{\cN=1}$ is generically much smaller as most of the scalars are stabilised at $\mino$.
Furthermore, we shall see that the role of the Goldstone bosons is the crucial difference between
the $\cN=1$ effective action arising from a spontaneously broken
$\cN=2$ theory  and that obtained by an $\cN=1$
truncation of the same $\cN=2$ theory
\cite{Andrianopoli:2001zh,Andrianopoli:2001gm,Andrianopoli:2002rm,Andrianopoli:2002vq,D'Auria:2005yg}
(see \cite{Grimm:2004uq,Grimm:2004ua,Benmachiche:2006df,Koerber:2007xk,Martucci:2009sf} for type II
orientifold compactification examples). The field space of the latter always
contains a submanifold of the $4 n_{\rm h}$-dimensional manifold
$\M_{\rm h}$ of maximal dimension $2n_{\rm h}$, rather than a quotient
of maximal dimension $4 n_{\rm h}-2$.

It is possible that the original $\cN=2$ supergravity
is also gauged with respect to Killing vectors which do not participate in the
partial supersymmetry breaking and which induce a separate mass scale $\mm$.
For $\mm > \mino$ all heavy multiplets with masses of ${\cal O}(\mm)$ should also be integrated out
and thus are not visible in the $\cN=1$ low-energy effective action.
If $\mm < \mino$, on the other hand, then the associated light multiplets
are kept in the action and do contribute to the superpotential ${\cal W}$
and possibly also to the D-terms ${\cal D}^{\hat I} $. Due to their $\cN=2$ origin, we will see that both ${\cal W}$ and ${\cal D}^{\hat I} $ take a special form.

The remainder of this paper is organised as follows. In Section~\ref{review} we briefly
summarise the results of \cite{Louis:2009xd} in order to set the
stage for our analysis. However, here we shall use a more geometric formulation of the hyperino supersymmetry conditions compared to \cite{Louis:2009xd}, stating them as a holomorphicity condition on the Killing vectors. In
Section~\ref{section:None} we then derive the $\cN=1$ low-energy
effective action. We begin with the target space metric of the scalar
fields in Section~\ref{section:NoneK}, show that it is K\"ahler and
determine its K\"ahler potential $K^{\cN=1}$. In
Section~\ref{section:Nonef} we compute the $\cN=1$
gauge kinetic function $f$
and check its holomorphicity with respect to the
$\cN=1$ complex structure.
Similarly, in Section~\ref{section:NoneW} we derive the
superpotential ${\cal W}$ and show its holomorphicity.  In Section~\ref{section:NoneD} we determine
the $D$-terms and in Section~\ref{section:SQC} we give the $\cN=1$ K\"ahler potential, the superpotential and the $D$-terms for the class of special quaternionic-K\"ahler manifolds.  We conclude in
Section~\ref{Conc}. In Appendix~\ref{section:massive_multi} we compute the normalised masses of
the two heavy gauge bosons and show their consistency with the
$\cN=1$ mass relations. In Appendix~\ref{section:holcoords} we show that the coordinates on the K\"ahler space introduced in Section~\ref{section:Min_special} are holomorphic.


\section{Partially broken $\cN=2$ supergravities}
\label{review}

\subsection{Gauged ${\cal N}=2$ supergravities
}
\label{section:N=2}

We shall first briefly recall the spectrum and couplings of four-dimensional ${\cal
N}=2$ supergravity (for a review see e.g.\ \cite{Andrianopoli:1996cm}). The theory consists of a gravitational multiplet, $\nv$ vector multiplets and $\nh$ hypermultiplets. The gravitational multiplet $(g_{\mu\nu},\Psi_{\mu {\cal A}}, A_\mu^0)$ contains the spacetime metric $ g_{\mu\nu}, \mu,\nu =0,\ldots,3$, two gravitini $\Psi_{\mu {\cal A}}, {\cal A}=1,2$, and the graviphoton $A_\mu^0$. A vector multiplet $(A_\mu,\lambda^{\cal A}, t)$ contains a vector $A_\mu$, two gaugini $\lambda^{\cal A}$  and a complex scalar $t$. Finally, a hypermultiplet $(\zeta_{\alpha}, q^u)$ contains two hyperini $\zeta_{\alpha}$ and 4 real scalars $q^u$. For $\nv$ vector- and $\nh$ hypermultiplets there are a total of $2\nv +4\nh$ real scalar fields and $2(\nv+\nh)$ spin-$\tfrac12$ fermions in the spectrum. For an ungauged theory the bosonic matter Lagrangian is given by
\begin{equation}\begin{aligned}\label{sigmaint}
{\cal L}\ =\  - \mathrm{i} \mathcal{N}_{IJ}\,F^{I +}_{\mu\nu}F^{\mu\nu\, J+}
+ \mathrm{i} \overline{\mathcal{N}}_{IJ}\,
F^{I-}_{\mu\nu} F^{\mu\nu\, J-}
+ g_{i\bar \jmath}(t,\bar t)\, \partial_\mu t^i \partial^\mu\bar t^{\bar \jmath}
+ h_{uv}(q)\, \partial_\mu q^u \partial^\mu q^v
\ ,
\end{aligned}\end{equation}
where  $h_{uv},\, u,v=1,\ldots,4\nh,$ is the metric on the
$4\nh$-dimensional space  ${\M}_{\rm h}$, which ${\cal N}=2$ supersymmetry constrains to be a quaternionic-K\"ahler manifold \cite{Bagger:1983tt,deWit:1984px}.  Such manifolds have a holonomy group given by $Sp(1)\times Sp(\nh)$. In addition, they admit a
triplet of complex structures $J^x, x=1,2,3$, which satisfy the quaternionic algebra
\begin{equation}\label{jrel}
J^x J^y = -\delta^{xy}{\bf 1} + \epsilon^{xyz} J^z ~.
\end{equation}
The metric $h_{uv}$ is Hermitian with respect to all
three complex structures. Correspondingly, a  quaternionic-K\"ahler manifold admits a triplet of hyper-K\"ahler two-forms given by $K^x_{uv} = h_{uw} (J^x)^w_v$ that are only covariantly closed with respect to the $Sp(1)$ connection $\omega^x$, i.e.
\begin{equation}\label{deriv_Sp(1)_curvature}
\nabla K^x \equiv dK^x + \epsilon^{xyz} \omega^y \wedge K^z=0 \ .
\end{equation}
In other words, $K^x$ is proportional to the $Sp(1)$ field strength of $\omega^x$, thus leading to
\begin{equation} \label{def_Sp(1)_curvature}
 K^x =\diff \omega^x + \tfrac12 \epsilon^{xyz} \omega^y\wedge \omega^z\ .
\end{equation}

The metric $g_{i\bar \jmath},\, i,\bar\jmath = 1,\ldots,\nv$, is defined on the $2\nv$-dimensional space ${\M}_{\rm v}$, which
${\cal N}=2$ supersymmetry constrains to be a  special-K\"ahler manifold \cite{deWit:1984pk,Craps:1997gp}. This implies that
the metric obeys
\begin{equation}\label{gdef}
g_{i\bar \jmath} = \partial_i \partial_{\bar \jmath} K^{\rm v}\ ,
\qquad \textrm{for}\qquad
K^{\rm v}= -\ln \iu \left( \bar X^I \cF_I - X^I\bar \cF_I \right)\ .
\end{equation}
Both $X^I(t)$ and ${\cal F}_I(t)$, $I= 0,1,\ldots,\nv$, are holomorphic functions of the scalars $t^i$ and in the ungauged case one can always choose $\cF_I = \partial\cF/\partial{X^I}$, i.e.\ $\cF_I$ is the derivative of a holomorphic prepotential $\cF(X)$  which is homogeneous of degree two. Furthermore, it is possible to go to a system of `special coordinates' where $X^I= (1,t^i)$ (See e.g. \cite{Craps:1997gp} for further details).

The $F^{I \pm}_{\mu\nu}$ that appear in
the Lagrangian \eqref{sigmaint} are the self-dual and anti-self-dual parts of
the usual field strengths. They include the field strengths of the gauge bosons of
the vector multiplets and the graviphoton.
Their kinetic matrix $
\mathcal{N}_{IJ}$ is a function of the $t^i $ given by
\begin{equation}
  \label{Ndef}
  {\cal N}_{IJ} = \bar \cF_{IJ} +2\iu\ \frac{\mbox{Im} \cF_{IK}\mbox{Im}
    \cF_{JL} X^K X^L}{\mbox{Im} \cF_{LK}  X^K X^L} \ ,
\end{equation}
where $\cF_{IJ}=\partial_I \cF_J$. As we shall discuss in Sections \ref{section:Nonef} and \ref{section:NoneD}, the second term in \eqref{Ndef} is due to the inclusion of the graviphoton in $F^{I \pm}_{\mu\nu}$.

In the ungauged case the equations of motion derived from $\cal L$
are invariant under $Sp(\nv+1)$ electric-magnetic duality rotations
which act on the $(2\nv+2)$-dimensional symplectic vectors $(F^I,
G_I)$ and $(X^I, \cF_I)$. The $G_I$ are dual magnetic field strengths that only appear on-shell, in that they are not part
of the Lagrangian \eqref{sigmaint} and are defined by
\begin{equation}\label{Gdualdef}
G_{I}^{\mu\nu\pm} = \pm \frac{\mathrm{i}}{2}\frac{\partial {\cal L}}{\partial F^{I\pm}_{\mu\nu}}~,
\end{equation}
from which we find (suppressing the spacetime indices)
\begin{equation}\label{Gdual}
G_I^+ =  \mathcal{N}_{IJ}\,F^{J +}\ , \qquad
G_I^- =  \overline{\mathcal{N}}_{IJ}\,F^{J -}\ . \qquad
\end{equation}

The symplectic invariance is broken in the presence of charged scalars,
i.e.\ in gauged supergravities,
and the resulting theory crucially depends on which charges (electric
or magnetic) the fermions and scalars carry. In fact,
one of the necessary conditions for partial supersymmetry breaking is
the appearance of magnetically charged fields
\cite{Antoniadis:1995vb,Ferrara:1995gu,Louis:2009xd}. Therefore, the
formalism of the embedding tensor introduced in
\cite{deWit:2002vt,deWit:2005ub} is ideally suited to discuss the
problem of partial supersymmetry breaking, as it treats the electric
vectors $A_\mu^{~~I}$ and their magnetic duals $B_{\mu I}$ on the
same footing and naturally allows for arbitrary gaugings.

As we shall review in the next section, partial supersymmetry breaking
needs at least two commuting
isometries in the hypermultiplet sector while it is sufficient for the vector multiplets
to be Abelian  \cite{Ferrara:1995gu,Louis:2009xd}.
Therefore, we focus on this situation and introduce covariant derivatives of the following form into the Lagrangian \eqref{sigmaint}:
\begin{equation}\label{d2}
\partial_\mu q^u\to D_{\mu} q^u
= \pt_{\mu}  q^u - A^{~I}_{\mu}\, \Theta_I^{~\lambda}\, \kk_{\lambda}^u +
B_{\mu I}\, \Theta^{I{\lambda}}\, \kk_{\lambda}^u \ ,
\end{equation}
where $\Theta$ is the embedding tensor and $\kk_\lambda(q)$ are the Killing
vectors on $\M_{\rm h}$. Mutual locality of electric and magnetic
charges additionally imposes
$\Theta^{I[{\lambda}} \Theta_{I}^{\phantom{I}\kappa]} = 0$. Inserting the replacement \eqref{d2} into the Lagrangian \eqref{sigmaint} introduces both electric and magnetic vector fields. This upsets the counting of degrees of freedom and leads to unwanted equations of motion. Therefore, the Lagrangian has to be carefully augmented by a set of two-form gauge potentials $B_{\mu\nu}^M$ with couplings that keep supersymmetry and gauge invariance intact. As we do not need these couplings in this paper, we refer the interested reader to the literature for further details~\cite{deWit:2002vt,deWit:2005ub,deVroome:2007zd}.

An analysis of the symplectic extension of the gauged $\cN=2$ supergravity Lagrangian in $D=4$ to include electric and magnetic charges has been carried out in \cite{Dall'Agata:2003yr,Sommovigo:2004vj,D'Auria:2004yi}. We are specifically interested in the scalar part of supersymmetry variations, i.e.\
\begin{equation}\label{susytrans2}\begin{aligned}
\delta_\epsilon \Psi_{\mu {\cal A}} ~=& ~ D_\mu \epsilon_{\cal A} - S_{\cal AB} \gamma_\mu \epsilon^{\cal B} + \ldots \, ,\nonumber\\
\delta_\epsilon \lambda^{i {\cal A}} ~=& ~ W^{i{\cal AB}}\epsilon_{\cal B}+\ldots \, ,\\
\delta_\epsilon \zeta_{\alpha} ~=& ~ N_\alpha^{\cal A} \epsilon_{\cal A}+\ldots \, ,\nonumber
\end{aligned}
\end{equation}
where the ellipses indicate further terms that vanish in a maximally symmetric ground state. The $\gamma_\mu$ are Dirac matrices and $\epsilon^{\cal A}$ is the $SU(2)$ doublet of spinors parametrising the $\cN=2$ supersymmetry transformations.\footnote{Note that the $SU(2)$ R-symmetry acts as the $Sp(1)$ introduced above on the quaternionic-K\"ahler manifold.}
$S_{\cal AB}$ is the mass matrix of the two gravitini, while $W^{i {\cal AB}}$ and $N_\alpha^{\cal A}$ are related to the mass matrices of the spin-$\tfrac12$ fermions. The symplectic extensions of these expressions in the embedding tensor formalism are given by
\begin{eqnarray}\label{susytrans3}\begin{aligned}
S_{\cal AB} ~=&~ \tfrac{1}{2} \e^{K^{\rm v}/2} {V}^\Lambda \Theta_\Lambda^{~\lambda} P_{\lambda}^x
(\sigma^x)_{\cal AB} \ ,\nonumber\\
W^{i{\cal AB}} ~=&~ \mathrm{i} \e^{K^{\rm v}/2} g^{i\bar \jmath}\,
(\nabla_{\bar \jmath}\bar {V}^\Lambda) \Theta_\Lambda^{~\lambda} P_{\lambda}^x (\sigma^x)^{\cal AB}
\ ,\\
N_\alpha^{\cal A} ~=&~ 2 \e^{K^{\rm v}/2} \bar {V}^\Lambda \Theta_\Lambda^{~\lambda} U^{\cal A}_{\alpha u} \kk^u_{\lambda}
\ ,\nonumber
\end{aligned}
\end{eqnarray}
where the
matrices $(\sigma^x)_{\cal AB}$ and $(\sigma^x)^{\cal AB}$ are found
by applying the $SU(2)$ metric $\varepsilon_{\cal{AB}}$ (and its
inverse) to the standard Pauli matrices $(\sigma^x)_{\cal A}^{~~\cal
B}$, $x=1,2,3$. From \eqref{d2} we see that the embedding tensor
$\Theta_\Lambda^{~~\lambda}$
has electric and magnetic components, which we combined in
\eqref{susytrans3} as
$\Theta_\Lambda^{~~\lambda} = (\Theta_I^{~~\lambda},-\Theta^{I\lambda})$.
Similarly,  ${V}^\Lambda$ is the holomorphic symplectic vector defined by ${V}^\Lambda = (X^I,{\cal F}_I)$ and its K\"ahler covariant derivative reads
$\nabla_i V^\Lambda = \partial_i V^\Lambda +K^{\rm v}_i V^\Lambda$, with $ K^{\rm v}_i = \partial_i K^{\rm v}$. ${\mathcal U}^{\mathcal A\alpha}_u $ is the vielbein on the quaternionic-K\"ahler manifold ${\M}_{\rm h}$ and is related to the metric $h_{uv}$ via
\begin{equation}\label{Udef}
h_{uv} = {\mathcal U}^{\mathcal A\alpha}_u \varepsilon_\mathcal{AB}
\mathcal C_{\alpha \beta} \mathcal U^{\mathcal B\beta}_v  \ ,
\end{equation}
where $\mathcal C_{\alpha \beta}$ is the $Sp(\nh)$ invariant metric. Finally, $P^x_\lambda$ is a triplet of Killing prepotentials defined by
\begin{equation}\label{Pdef}
- 2 \kk^u_\lambda\,K_{uv}^x =   \nabla_v P_\lambda^x = \partial_v P_\lambda^x + \epsilon^{xyz} \omega^y_v P_\lambda^z  \ ,
\end{equation}
where $k^u_\lambda$ are
the isometries on the quaternionic-K\"ahler manifold and $K_{uv}^x$ is
the triplet of hyper-K\"ahler two-forms.

\subsection{Partial supersymmetry breaking}\label{section:vectors}

Spontaneous $\cN=2\to \cN=1$ supersymmetry breaking in a Minkowski or anti-de Sitter (AdS) ground state requires that for one linear combination
of the two spinors $\epsilon^{\cal A}$ parametrising the supersymmetry transformations, say $\epsilon^{\cal A}_1$, the variations of the fermions given in \eqref{susytrans2} vanish, i.e.\ $\delta_{\epsilon_1} \lambda^{i {\cal A}} = \delta_{\epsilon_1} \zeta_\alpha = \delta_{\epsilon_1} \Psi_{\mu {\cal A}} =0$.
Using the fact that in a supersymmetric Minkowski or AdS background the supersymmetry parameter obeys the Killing spinor equation\footnote{Note that the index of $\epsilon^*_{1\,\cal A}$ is not lowered with $\varepsilon_{\cal AB}$ but $\epsilon^*_{1\,\cal A}$ is related to $\epsilon^A_1$ just by complex conjugation. $|\mu|$ is related to the cosmological constant via $\Lambda = - 3 |\mu|^2$, while the phase of $\mu$ is unphysical.}
\begin{equation}
 D_\nu \epsilon_{1\,\cal A} = \tfrac12 \mu \gamma_\nu
 \epsilon^*_{1\,\cal A} \ ,
\end{equation}
the supersymmetry variations \eqref{susytrans2} yield
\begin{equation} \label{N=1conditions}
W_{i\cal AB}\, \epsilon^{\cal B}_1 ~= 0~ =~ N_{\alpha \cal A}\,
\epsilon^{\cal A}_1 \qquad \textrm{and}\qquad  S_\mathcal{AB}\,
\epsilon^{\cal B}_1~ =~  \tfrac12 \mu \epsilon^*_{1\,\cal A} \ .
\end{equation}
The second, broken generator, denoted by $\epsilon^{\cal A}_2$, should obey
\begin{equation} \label{N=1conditions2}
W_{i\cal AB}\, \epsilon^{\cal B}_2 \neq 0\qquad \textrm{or}\qquad
N_{\alpha \cal A}\, \epsilon^{\cal A}_2\neq 0\qquad \textrm{and}\qquad  S_\mathcal{AB}\,
\epsilon^{\cal B}_2~ \ne~  \tfrac12 \mu' \epsilon^*_{2\,\cal A} \ ,
\end{equation}
for any $\mu'$ that obeys $|\mu'|=|\mu|$, i.e.\ $\mu'$ only differs from $\mu$ by an unphysical phase.

A necessary condition for the existence of an $\cN=1$ ground state is that the two eigenvalues $m_{\Psi_1}$ and $m_{\Psi_2}$ of the gravitino mass matrix $S_{\cal AB}$ are non-degenerate, e.g.\ $m_{\Psi_1}\neq m_{\Psi_2}$. One of the two gravitini has to remain massless, i.e. $m_{\Psi_1}=0$ in a Minkowski ground state, while the second one becomes massive. The unbroken $\cN=1$ supersymmetry also implies that the massive gravitino has to be a member of an entire $\cN=1$ massive spin-$3/2$ multiplet, which has the spin content $s=(3/2,1,1,1/2)$. This means that two vectors, say $A_\mu^1, A_\mu^2$ and a spin-$1/2$ fermion $\chi$ have to become massive, in addition to the gravitino.\footnote{In Appendix \ref{section:massive_multi} we explicitly check that the correct $\cN=1$ mass relations are obeyed using the results of Section \ref{section:None}} Therefore, the would-be Goldstone fermion (the Goldstino), which gets eaten by the gravitino, is accompanied by two would-be Goldstone bosons (the sGoldstinos) that are eaten by the vectors \cite{Ferrara:1983gn}. In the resulting Lagrangian, only $\cN=1$ supersymmetry is linearly realized while the second, spontaneously broken supersymmetry generator acts non-linearly on the fields. When integrating out the massive fields, the latter is broken explicitly and we end up with an $\cN=1$ effective action.

The sGoldstinos necessarily arise from the hypermultiplets,
which means that ${\M}_{\rm h}$ has to admit at least two
commuting isometries, say $k_1$ and $k_2$, and that these isometries
have to be gauged \cite{Ferrara:1995gu,Fre:1996js}.
The corresponding Goldstone bosons are then charged and generate
the masses for the two heavy gauge bosons via the Higgs mechanism.
If  ${\M}_{\rm h}$ has further Killing vectors $k_\lambda,
\lambda\neq1,2$, which are gauged, then
additional charged and possibly massive scalars arise.
In fact, in \cite{Louis:2009xd} we showed
that only two Killing vectors can participate in the partial
supersymmetry breaking.  The other, orthogonal Killing vectors
either preserve the full $\cN=2$ supersymmetry, as analysed in \cite{Hristov:2009uj},
or break it completely.
In the latter case we need to assume
that this breaking is at a scale far below $m_{3/2}$ and therefore
can be neglected in the following discussion. However, we shall return to this
issue in Sections
\ref{section:NoneW} and \ref{section:NoneD} where we compute the
$\cN=1$ effective potential generated by such additional Killing vectors.

The definition \eqref{Pdef}
implies that the two non-trivial Killing vectors have
non-zero Killing prepotentials
$P_1^x, P_2^x$ in the $\cN=1$ background. For an $\cN=1$ solution these prepotentials
must not be proportional to each other, as this would allow us to take
linear combinations of $k_1$ and $k_2$ such that one combination
has vanishing prepotentials.
However, we can use the local $SU(2)$ invariance of the
hypermultiplet sector to rotate into a
convenient $SU(2)$-frame where
$P_{1,2}^x$
both lie entirely in
the $(x=1,2)$-plane. Thus, without loss of generality we can arrange
\begin{equation}\label{P3}
 P^3_1 ~ =~ P^3_2 ~ =~ 0~ =~ \partial_u P^3_1 ~ = ~ \partial_u P^3_2\ .
\end{equation}

From \eqref{susytrans3} we learn that in such a frame both
$S_{\cal  AB}$ and $W^{i{\cal AB}}$ are diagonal
in $SU(2)$ space and hence one can further choose  the parameter of
the unbroken $\cN=1$ generator to be $\epsilon_1 = {\epsilon\choose 0}$
or $\epsilon_1 = {0\choose\epsilon}$. This corresponds to the choice of
$\Psi_{\mu\, 1}$ or $\Psi_{\mu\, 2}$ as the massless
$\cN=1$ gravitino.\footnote{Note that all our expressions can also
be written in an $SU(2)$-covariant way by replacing the
``$3$''-direction with $\epsilon_1^A \sigma^x_{\cal AB} \epsilon_2^B$
and the direction spanned by $(P^1-\iu P^2)$ with $\epsilon_1^A
\sigma^x_{\cal AB} \epsilon_1^B$. So, for instance, \eqref{P3} then
reads
$\epsilon_1^A \sigma^x_{\cal AB} \epsilon_2^B P^x_{1,2} = \epsilon_1^A \sigma^x_{\cal AB} \epsilon_2^B \diff P^x_{1,2}=0.
$
}

After these preliminaries, let us now review the conditions for partial supersymmetry breaking
which we derived in \cite{Louis:2009xd}.

\subsubsection{Gravitino and gaugino equations}

For $\epsilon_1 = {\epsilon\choose 0}$
 the $\cN=1$ solution of the gravitino and gaugino
variations in a Minkowski vacuum was found to be \cite{Louis:2009xd}
\begin{equation}\label{solution_embedding_tensor}
\begin{aligned}
\Theta_I^{\phantom{I}1} =  -\Im\left(P^+_2\, {\cal F}_{IJ}\,\C^J \right) \ ,& \qquad  \Theta^{I1} =  -\Im \left(P^+_2\,\C^I\right) \ , \\
\Theta_I^{\phantom{I}2} =  \quad \Im\left(P^+_1\, {\cal F}_{IJ}\,\C^J \right) \ ,& \qquad  \Theta^{I2} =  \quad \Im \left(P^+_1\, \C^I\right) \ ,
\end{aligned}
\end{equation}
parametrised in terms of a complex vector $\C^I$. The mutual locality constraint then demands
\begin{equation} \label{constraint_embedding_tensor}
  \bar{\C}^{I} (\Im {\cal F})_{IJ}\, \C^{J}  = 0 \ ,
\end{equation}
and we have defined
\begin{equation}\label{Pc}
 P_{1,2}^\pm = P_{1,2}^1 \pm \iu P_{1,2}^2 \ .
\end{equation}
Note that the $\cN=1$ solution \eqref{solution_embedding_tensor} determines the embedding tensor in terms of $C^I$ but does not constrain the special-K\"ahler manifold ${\M}_{\rm v}$.

For $\epsilon_1 = {\epsilon\choose 0}$  the $\cN=1$ solution of the gravitino and gaugino
variations in an AdS vacuum was found to be \cite{Louis:2009xd}
\begin{equation}\label{solution_embedding_tensor_AdS}
\begin{aligned}
\Theta_I^{\phantom{I}1} = & -\Im\left({\cal F}_{IJ}\,(P^+_2\, \C_{\rm AdS}^J + \e^{K^{\rm v}/2} \tfrac{\bar \mu}{P^+_1} X^J) \right) \ , \\
\Theta^{I1} = & -\Im \left(P^+_2\,\C_{\rm AdS}^I + \e^{K^{\rm v}/2} \tfrac{\bar \mu}{P^+_1} X^I\right) \ , \\
\Theta_I^{\phantom{I}2} = & \quad \Im\left({\cal F}_{IJ}\,(P^+_1\, \C_{\rm AdS}^J - \e^{K^{\rm v}/2} \tfrac{\bar \mu}{P^+_2} X^J) \right) \ , \\
\Theta^{I2} = & \quad \Im \left(P^+_1\, \C_{\rm AdS}^I- \e^{K^{\rm v}/2} \tfrac{\bar \mu}{P^+_2} X^I\right) \ ,
\end{aligned}
\end{equation}
where again $\C_{\rm AdS}^I$ is a complex vector. The mutual locality constraint \eqref{constraint_embedding_tensor} now reads
\begin{equation}\label{constraint_embedding_tensor_AdS}
 \bar{\C}_{\rm AdS}^{I} (\Im {\cal F})_{IJ} \C_{\rm AdS}^{J} = - \tfrac{|\mu|^2}{2 |P_1|^2 |P_2|^2} \ .
\end{equation}


\subsubsection{Hyperino equations}\label{section:hyperino}
The solution to the hyperino equations is more model
dependent. We already stated that the  quaternionic-K\"ahler manifold
${\M}_{\rm h}$ has to  admit two commuting isometries with Killing
prepotentials $P^x_1$ and $P^x_2$ that are not proportional to each
other in the $\cN=1$ locus. In addition, the $\cN=1$ hyperino supersymmetry conditions
\begin{equation} \label{Ncond}
N_{\alpha \cal A}\, \epsilon_1^{\cal A}~ =~ N_{\alpha 1}~ =~ 0
\end{equation}
have to be satisfied. Before we continue, let us rewrite \eqref{Ncond} in a more convenient form.
The insertion of \eqref{susytrans3} into \eqref{Ncond} and subsequent complex conjugation implies
\begin{equation}\label{kcond}
 k^u\, {\mathcal U}_{\alpha u}^{2}\ =\ 0\ ,
\end{equation}
where we have defined
\begin{equation}\label{knew}
 k^u =  {V}^\Lambda \big(\Theta_\Lambda^{~~1} \kk_{1}^u +\Theta_\Lambda^{~~2} \kk_{2}^u\big)\ .
\end{equation}
By contracting the decomposition \cite{Andrianopoli:1996cm,D'Auria:2001kv}
\begin{equation}\label{Udecomp}
{\mathcal U}_{\alpha u}^{\mathcal A}{\mathcal U}_v^{\mathcal B\alpha}
= - \tfrac{\iu}{2} K^x_{uv}\sigma^{x {\mathcal A}{\mathcal B}} - \tfrac12 h_{uv}
\epsilon^{{\mathcal A}{\mathcal B}}\ ,
\end{equation}
with $k^v$ and using the explicit expressions 
\begin{equation}
(\sigma^1)^{\cal AB} =  \left(\begin{array}{cc}-1 &0\\ 0& 1 \end{array}\right)~, \ (\sigma^2)^{\cal AB} =  \left(\begin{array}{cc} -\mathrm{i} &0\\ 0& -\mathrm{i} \end{array}\right)~ , \ (\sigma^3)^{\cal AB} =  \left(\begin{array}{cc}0 &1\\ 1& 0 \end{array}\right)~,
\end{equation}
we see that \eqref{kcond} is equivalent to
\begin{equation}\label{jhol}
k^u \left(J^{1~v}_{~~u} - \iu J^{2~v}_{~~u}\right) = 0 \ , \qquad
k^u J^{3~v}_{~~u} = \iu k^{v}\ .
\end{equation}
The second condition of \eqref{jhol} simply states that
$k$ is holomorphic with respect to the complex structure $J^3$. Furthermore, using the relation between the three $J$'s given in \eqref{jrel}, the first equation in \eqref{jhol} follows from the second one.
For our subsequent analysis it is convenient to
define a new pair of Killing vectors $k^u_{1,2}$
by using the real and imaginary parts of the $k^u$ defined in \eqref{knew}, such that the following holds\footnote{In order to keep the notation simple we shall use the same letter $k$ to denote the original Killing vectors, as well as the redefined ones. The same holds for the respective Killing prepotentials $P^x$.}
\begin{equation} \label{Jk}
J^{3~v}_{~~u} k_1^u = -k_2^v \ , \qquad J^{3~v}_{~~u} k_2^u = k_1^v \ .
\end{equation}
Note that this is nothing more than a change of basis in the space spanned by the two Killing vectors. The coefficients in this change of basis do not depend on the coordinates of $\M_{\rm h}$, as the embedding tensor components are constant. As the related Killing prepotentials $P^x_{1,2}$ will also not be proportional to each other, we can equally use the new Killing vectors to construct a partial supersymmetry breaking solution, instead of the original Killing vectors
$\kk_{1,2}$ appearing in \eqref{d2}.

The conditions \eqref{jhol}, or equivalently \eqref{Jk}, also constrain the Killing prepotentials. Written in terms of the associated K\"ahler forms the first condition of \eqref{jhol} reads
\begin{equation}\label{relationkK}
 k_1^uK^1_{uv}=-k_2^uK^2_{uv} \ , \qquad k_1^uK^2_{uv}=k_2^uK^1_{uv} \ ,
\end{equation}
which, together with the definition of the prepotentials \eqref{Pdef}, implies
\begin{equation}
 P_1^1=-P_2^2 \ , \qquad P_1^2=P_2^1 \ .
\end{equation}
This in turn simplifies the embedding tensor solutions \eqref{solution_embedding_tensor}, which after a redefinition of $C^I$ read
\begin{equation}\label{solution_embedding_tensor2}
\begin{aligned}
\Theta_I^{\phantom{I}1} = & \Re\big( {\cal F}_{IJ}\,\C^J \big) \ , \qquad  \Theta^{I1} = & \Re \C^I \ , \\
\Theta_I^{\phantom{I}2} = & \Im\big( {\cal F}_{IJ}\,\C^J \big) \ , \qquad  \Theta^{I2} = & \Im  \C^I \ .
\end{aligned}
\end{equation}
Similarly, the AdS solutions \eqref{solution_embedding_tensor_AdS} become
\begin{equation}\label{solution_embedding_tensor_AdS2}
\begin{aligned}
\Theta_I^{\phantom{I}1} = & \Re\big({\cal F}_{IJ}\,( \C_{\rm AdS}^J - \iu \e^{K^{\rm v}/2} \tfrac{\bar \mu}{P^+_1} X^J) \big) \ , \\
\Theta^{I1} = & \Re \big(\C_{\rm AdS}^I - \iu  \e^{K^{\rm v}/2} \tfrac{\bar \mu}{P^+_1} X^I\big) \ , \\
\Theta_I^{\phantom{I}2} = &  \Im\big({\cal F}_{IJ}\,(\C_{\rm AdS}^J +\iu \e^{K^{\rm v}/2} \tfrac{\bar \mu}{P^+_1} X^J) \big) \ , \\
\Theta^{I2} = &  \Im \big(\C_{\rm AdS}^I + \iu \e^{K^{\rm v}/2} \tfrac{\bar \mu}{P^+_1} X^I\big) \ .
\end{aligned}
\end{equation}

The hyperino conditions \eqref{Jk}, or equivalently \eqref{Ncond}, are difficult to solve in general. In \cite{Louis:2009xd} we showed that for special quaternionic-K\"ahler manifolds, i.e.\ quaternionic-K\"ahler manifolds that are in the image of the c-map \cite{Cecotti:1988qn}, \eqref{Ncond} together with all other constraints can be fulfilled.\footnote{Explicit examples of AdS vacua are constructed in \cite{Lust:2004ig,House:2005yc,Micu:2006ey,Tomasiello:2007eq,KashaniPoor:2007tr,Cassani:2009na,Cassani:2009ck,Lust:2009mb}.} In the following, however, we do not restrict our analysis to this class of manifolds but instead only assume that an $\cN=1$ solution exists, i.e.\
we assume that equations ~\eqref{Jk}, \eqref{solution_embedding_tensor2} and \eqref{solution_embedding_tensor_AdS2} are satisfied without specifying a particular explicit solution.

Before we continue let us note that the $\cN=1$ solution we just
recalled has both $W_{i\cal AB}\, \epsilon^{\cal B}_2 \neq 0$  \emph{and}
$N_{\alpha \cal A}\, \epsilon^{\cal A}_2\neq 0$. In
\eqref{N=1conditions2}
we allowed for the logical possibility that supersymmetry is only broken in
the gaugino or hyperino sector. However, this situation cannot
occur for partial supersymmetry breaking.
The two Killing prepotentials $P^x_{1,2}$ have to be non-zero in
order to render the two eigenvalues of the gravitino mass matrix
$S_{\cal AB}$ non-degenerate. Using \eqref{Pdef} or the equivariance condition $  2 k_1^u k_2^v K^x_{uv} + \epsilon^{xyz} P^y_1 P^z_2 = 0$ \cite{Andrianopoli:1996cm}, we can further
conclude that the two Killing vectors  $k^u_{1,2}$
have to be non-zero
which, together with \eqref{susytrans3}, implies  $N_{\alpha \cal A}\neq
0$.
Finally, one can check that for the charges \eqref{solution_embedding_tensor2} and \eqref{solution_embedding_tensor_AdS2} $W_{i\cal AB}$ is always non-zero.

\subsubsection{Massive, light and massless scalars}\label{section:scales}

The Minkowski and AdS ground states described above are local $\cN=1$
minima in $\cN=2$ field space i.e.\ the $\cN=2$ supersymmetry
variations were solved for an $\cN=1$ vacuum which can be a point in
each of ${\M}_{\rm  h}$ and ${\M}_{\rm  v}$ or a higher-dimensional
vacuum manifold. In the latter case there are
exactly flat directions (moduli) of the minimum along which $\cN=1$
supersymmetry is preserved.
In addition, there can be light scalars in the spectrum
(i.e.\ with masses $\mm$ much smaller than $\mino$)
which either preserve $\cN=1$ supersymmetry or break it
at a scale beneath $\mino$. This breaking is negligible in the limit
$\mm\ll \mino$ and therefore we also include \emph{all} light scalar fields
in the definition of the $\cN=1$ field space.
As we will see in Sections \ref{section:NoneW} and
\ref{section:NoneD} the light fields contribute to the
superpotential and D-terms in the effective action  and any spontaneous
$\cN=1$ supersymmetry breaking will be captured by these couplings.
In the following we
denote the scalars of the $\cN=1$ field space by $\hat{t}$ and
$\hat{q}$, where there is natural split into fields descending from
the $\cN=2$ vector- and hypermultiplets, respectively.

Let us now give a more precise description of the distinction between
scalars with masses of ${\cal O}(\mino)$ and massless (or light)
scalar fields.  The latter are the deformations which preserve the
$\cN=1$ supersymmetry conditions \eqref{N=1conditions} in the limit
$\mm\to0$. Equivalently, \eqref{jhol} holds and the embedding tensor solutions \eqref{solution_embedding_tensor2} or
\eqref{solution_embedding_tensor_AdS2} remain constant across the
$\cN=1$ field space. On the other hand, any deformation that violates
the $\cN=1$ supersymmetry conditions \eqref{N=1conditions}
(ignoring any supersymmetry breaking at a lower scale $\mm$)
should have
a mass of ${\cal O} (\mino)$. Consistency of the low-energy effective
theory implies that all fields with a mass of ${\cal O} (\mino)$ should
be integrated-out along with the massive gravitino.

As an example, let us consider the Minkowski solution
\eqref{solution_embedding_tensor2} at a point $t=t_0$ and determine
the deformations $t=t_0+\delta t$ which preserve
\eqref{solution_embedding_tensor2}.
This implies
\begin{equation}\label{deformV}
{\cal F}_{IJK} C^J \delta X^K = 0 \ .
\end{equation}
For a generic prepotential ${\cal F}$, \eqref{deformV} gives
$n_\textrm{v}$ equations for $n_\textrm{v}$ deformation
parameters. This can be seen by noting that the homogeneity of the holomorphic prepotential ${\cal F}$ implies ${\cal F}_{IJK}X^K = 0$. Thus all $n_\textrm{v}$ scalars in the vector
multiplets are generically stabilised with masses of ${\cal O}(\mino)$
and an $\cN=1$ moduli space can only occur for special prepotentials . For example,
if the prepotential ${\cal F}$ is purely quadratic, \eqref{deformV} is satisfied
on the entire field space and no scalars in the vector multiplets are stabilised. This corresponds to $\M_{\rm v}= SU(1,n_{\rm v})/SU(n_{\rm v})$. In contrast, for a generic cubic prepotential \eqref{deformV} tells us that all scalars are stabilised. This would appear to be in conflict with the existence of the $n_\textrm{v}$ shift isometries on $\M_{\rm v}$ \cite{deWit:1992wf}. However, these shift isometries induce symplectic rotations on the vectors of the theory. These symplectic rotations are only symmetries of the ungauged theory and can be broken by the charges $\Theta_{\Lambda}^{1,2}$ given in \eqref{solution_embedding_tensor2}. The same conclusion can be reached for isometries on general special-K\"ahler manifolds.

A computation analogous to \eqref{deformV} for the AdS solution \eqref{solution_embedding_tensor_AdS2} leads to
\begin{equation}\label{deformV_AdS}
{\cal F}_{IJK} C^J \delta X^K + 2 \tfrac{\mu}{P^-_1} (\Im {\cal F})_{IJ} \delta (\e^{K^{\rm v}/2} \bar X^J) = 0 \ .
\end{equation}
In contrast to the Minkowski case, this is not a holomorphic equation. Nevertheless the number of equations coincides with the number of scalars in the vector multiplets and generically all scalars are stabilised.

A corresponding condition arises for the scalars of ${\M}_{\rm h}$ from \eqref{Ncond} or equivalently
\eqref{Jk}. The Killing vector $k=k_1 + \iu k_2$ should stay holomorphic over the
entire $\cN=1$ field space or in other words
\begin{equation}\label{deformH}
\delta \, ( J^{3~v}_{~~u} k_1^u + k_2^{v})\ = 0
\end{equation}
should hold. This condition generically stabilises a large number of scalar fields arising
from the hypermultiplet sector. In contrast to the vector multiplet sector, a non-trivial $\cN=1$ moduli
space necessarily arises whenever ${\M}_{\rm h}$ has additional
isometries which commute with the two isometries responsible for the partial
supersymmetry breaking. We will return to this issue in Section~\ref{section:SQC}.


\section{The low-energy effective $\cN=1$ theory}\label{section:None}
\setcounter{equation}{0}

Let us now turn to the main objective of
this paper and derive the low-energy effective $\cN=1$ theory that is
valid below the scale of supersymmetry breaking set by $\mino$.
We will begin by outlining the procedure employed and
briefly summarising the results which we obtain.

In the previous section we reviewed the properties of an $\cN=2$
supergravity that admits $\cN=1$ Minkowski or AdS backgrounds. Consistency requires that an $\cN=1$ massive spin-3/2 multiplet with spins $s=(3/2,1,1,1/2)$ and mass $\mino$ is generated, possibly along with a set of massive $\cN=1$ chiral- and vector multiplets whose masses are also of ${\cal O}(\mino)$. All of these multiplets  have to be integrated out to obtain the $\cN=1$ low-energy effective action.\footnote{If the $\cN=2$ theory has a supersymmetric mass scale above $\mino$ then all multiplets at that scale are also integrated out.}  At the two-derivative level this is achieved by using the equations of motion of the massive fields to first non-trivial order in $p/\mino$, where $p\ll\mino$ is the characteristic momentum.  The low-energy effective theory should then contain the leftover light $\cN=1$ multiplets, i.e.\ the gravity multiplet, $n'_{\rm v}$ vector multiplets and $n_{\rm c}$ chiral multiplets. These multiplets either have a mass  below $\mino$ or are exactly massless. The case when all the multiplets are massless arises when the $\cN=2$ supergravity is gauged with respect to just the two Killing vectors that are responsible for the partial supersymmetry breaking.  If, on the other hand, the $\cN=2$ supergravity is gauged with respect to additional Killing vectors, then some of the $\cN=1$ multiplets can have a light mass or, more generally, contribute to the $\cN=1$ effective potential. However, the derivation of the low-energy effective action is insensitive additional gaugings. Whether or not such gaugings preserve the $\cN=1$ supersymmetry or spontaneously break it only becomes clear on examining the ground states of the effective potential.

Integrating out all massive fields of ${\cal O}(\mino)$ in the $\cN=2$
gauged supergravity should naturally lead to an $\cN=1$ effective
theory. Its bosonic matter Lagrangian therefore has a standard form, given by
\cite{Wess:1992cp,Gates:1983nr}
\begin{eqnarray}\label{N=1Lagrangian}
  \hat{\cal L} ~=~ - \ K_{\hat A  \hat{\bar B} } D_\mu M^{\hat A} D^{\mu} \bar M^{\hat{\bar B} } - \tfrac{1}{2} f_{\hat I \hat J}\ F^{\hat I -}_{\mu\nu}F^{\mu\nu\, \hat J -} - \tfrac{1}{2}\bar{f}_{\hat I \hat J}\ F^{\hat I +}_{\mu\nu} F_{\rho\sigma}^{\hat J + } - V \ ,
\end{eqnarray}
where
\begin{eqnarray}\label{N=1pot}
  V ~=~ V_F + V_{\cal D} ~=~
  e^K \big( K^{\hat A \hat{\bar B}} D_{\hat A} {\cal W} {D_{\hat{\bar B}} \bar {\cal W}}-3|{\cal W}|^2 \big)
  +\tfrac{1}{2}\,
  (\text{Re}\; f)_{\hat I \hat J} {\cal D}^{\hat I} {\cal D}^{\hat J}
  \ .
\end{eqnarray}
We use hatted indices to label the fields of the $\cN=1$
effective theory. $M^{\hat A} = M^{\hat A} (\hat{t},\hat{q}) $ collectively denotes
all complex scalars in the theory, i.e.\ those descending from both the
vector- and hypermultiplet sectors in the original ${\cal N}=2$ theory.
$K_{\hat A \hat{\bar B} } $ is a K\"ahler metric satisfying $ K_{\hat
  A \hat{\bar B} } = \partial_{\hat A} \bar\partial_{\hat{\bar B}}
K(M,\bar M)$. $F^{\hat I +}_{\mu\nu}$ and $F^{\hat I -}_{\mu\nu}$ denote the self-dual and
anti-self-dual $\cN=1$ gauge field strengths, respectively, and $f_{\hat I \hat J}$
is the holomorphic gauge kinetic function. The scalar potential $V$ is determined in terms of the holomorphic
superpotential ${\cal W}$, its K\"ahler-covariant derivative $D_{\hat A}
{\cal W}=
\partial_{\hat A} {\cal W} + (\partial_{\hat A} K)\, {\cal W}$ and the
D-terms ${\cal D}^{\hat I}$, given by
\begin{equation}\label{NoneDterm}
  {\cal D}^{\hat I}~ =~ -2\, (\text{Re}\; f)^{-1 \hat I \hat J}\, {\cal P}_{\hat J}~,
\end{equation}
where ${\cal P}_{\hat J}$ is the $\cN=1$ Killing prepotential.

The objective of this section is to compute the coupling
functions $K,{\cal W},f$ and ${\cal P}$ of the effective $\cN=1$
theory in terms of $\cN=2$ `input data'. $\cN=1$ supersymmetry
constrains $\cal W$ and $f$ to be holomorphic while the metric $K_{\hat A
  \hat{\bar B} }$ has to be K\"ahler. Showing that the low-energy
effective theory has these properties serves as an important
consistency check of our results.

Before we turn to the derivation of these couplings let us briefly anticipate the results. One interesting aspect relates to the $\cN=1$ scalar manifold that descends from the $\cN=2$ product space
$\M = {\M}_{\rm h}\times {\M}_{\rm v}$, where ${\M}_{\rm v}$ is
already a K\"ahler manifold but ${\M}_{\rm h}$ is not. In
Section~\ref{section:NoneK} we will show that integrating out the two
heavy gauge bosons in the gravitino multiplet amounts to taking a
quotient of ${\M}_{\rm h}$ with respect to the two gauged isometries
$k_1,k_2$ discussed in the previous section. This quotient, denoted by
\begin{equation}\label{quotient}
{\Mq}_{\rm h} = {\M}_{\rm    h}/\langle k_1,k_2 \rangle\ ,
\end{equation}
has co-dimension two, corresponding to the fact that the two Goldstone
bosons giving mass to the two gauge bosons have been removed. We
shall see that the quotient $\Mq_{\rm h}$ is indeed
K\"ahler, which establishes the consistency with $\cN=1$ supersymmetry.
In order to obtain the final $\cN=1$ scalar field space, we also have to
integrate out all additional scalars that gained a mass of ${\cal O}
(\mino)$. However, these scalars are not Goldstone bosons and thus
integrating them out corresponds to simply projecting ${\M}_{\rm
  v}\times {\Mq}_{\rm h}$ to a K\"ahler subspace $\M^{\cN=1} = {\Mq}_{\rm  v} \times {\Mq}_{\rm  h}\ ,$
where ${\Mq}_{\rm v}$ coincides with ${\M}_{\rm v}$ or is a
submanifold thereof. ${\Mq}_{\rm h}$ can also be a subspace of \eqref{quotient}, but for notational simplicity we do not introduce a separate symbol for this.

Integrating out the two massive gauge bosons projects the
$\cN=2$ gauge kinetic function to a submatrix. In
Section~\ref{section:Nonef} we will show that one of the two massive gauge
bosons is always given by the graviphoton.\footnote{This can also be seen by noting that  \eqref{constraint_embedding_tensor} implies that $C^I$ consists of a spacelike and a timelike component with respect to $\Im{\cal F}_{IJ}$, which has signature $(1,n_{\rm v})$. The timelike component corresponds to a gauging with respect to the graviphoton.} Integrating out this vector
leads to a holomorphic gauge kinetic function $f$ that is the second
derivative of the holomorphic prepotential on ${\Mq}_{\rm v}$, similarly to the case of $\cN=1$ truncations \cite{Andrianopoli:2001zh,Andrianopoli:2001gm}.

Finally, as our $\cN=1$ effective theory descends from an $\cN=2$ supergravity, its superpotential ${\cal W}$ and the D-terms can only be non-trivial if there are additional charged scalars present, i.e.\ if there are further gaugings at a scale beneath $\mino$. As discussed above, this precisely occurs when isometries other than $k_1$ and $k_2$ are gauged in the original $\cN=2$ theory.
Since both ${\cal W}$ and ${\cal D}$ appear in the $\cN=1$ supersymmetry
transformations of the gravitino and gaugini, we can consider the corresponding $\cN=2$
supersymmetry transformations restricted to $\cN=1$ fields and then read off the appropriate terms. We will carry this out in Sections
\ref{section:NoneW} and \ref{section:NoneD}. Using the complex structure of $\M^{\cN=1}$, we will then also check the holomorphicity of ${\cal W}$ in Section~\ref{section:NoneW}.

Let us now turn to the detailed derivation of the $\cN=1$ couplings,
starting with the metric on the quotient
${\Mq}_{\rm h}$.

\subsection{The K\"ahler metric on the quotient ${\Mq}_{\rm h}$}
\label{section:NoneK}

The first step in determining the sigma-model metric on the quotient
${\Mq}_{\rm h}$ is to eliminate the two massive gauge bosons via their
field equations, which are algebraic in the limit $p\ll\mino$.  In
order to be able to use the constraints \eqref{jhol} and \eqref{Jk} derived from the hyperino conditions, we
first have to rewrite the combination $\Theta_\Lambda^\lambda\kk_\lambda,~ \lambda=1,2,$
that appears in \eqref{d2} in terms of the new
Killing vectors defined in \eqref{knew}.  This change of basis can be
compensated by an appropriate change of $\Theta_\Lambda^\lambda$, such that the
covariant derivatives given in \eqref{d2} continue to have the same
form, albeit with rotated $\kk_\lambda$ and $\Theta_\Lambda^\lambda$ (for simplicity, we shall not introduce new symbols for the rotated quantities).
{}From \eqref{sigmaint} we then obtain
\begin{equation}\label{eofA}
  \frac{\partial \cal L}{\partial A_\mu^{\lambda}} =
  -2 k^v_{\lambda} h_{uv} \partial_{\mu} q^u
  + m_{\lambda\rho}^2 A_\mu^{\rho} = 0\ ,\qquad \lambda, \rho =1,2\ ,
\end{equation}
where we have defined
\begin{equation}\label{Acomb}
  A_\mu^\lambda \equiv A_\mu^\Lambda\Theta^\lambda_\Lambda =
  A_\mu^I \Theta^\lambda_I -B_{\mu I}\Theta^{I\lambda}\ ,
\end{equation}
and the mass matrix
\begin{equation}\label{mdef}
  m_{\lambda\rho}^2 = 2 k^u_{\lambda} h_{uv} k^v_{\rho}\ .
\end{equation}
Using the quaternionic algebra \eqref{jrel} and the hyperino conditions \eqref{Jk} written in terms of the associated
K\"ahler forms $K^x$, we see that this mass matrix is diagonal
\begin{equation}
  m_{\lambda\rho}^2 = m^2 \, \delta_{\lambda\rho}  \ ,
\end{equation}
where
\begin{equation}\label{massrel}
  m^2 = 2 |k_{1}|^2 = 2 |k_{2}|^2 \ .
\end{equation}
Inserting the algebraic field equations \eqref{eofA} back into the
Lagrangian yields a modified kinetic term for the hypermultiplet
scalars, which reads
\begin{equation}
  \hat{\cal L} = \h_{uv} \partial_{\mu} q^u \partial^{\mu} q^v\ .
\end{equation}
$\h_{uv}$ is the metric on the quotient ${\Mq}_{\rm h}$
and is given by
\begin{equation}\label{hq}
  \h_{uv} = 
  h_{uv} - \frac{2k_{1 u} k_{1 v} + 2k_{2 u} k_{2 v}}{m^{2}} = \PQ^w_u h_{wv}   \ ,
\end{equation}
where
$k_{\lambda u} = k^w_{\lambda} h_{wu}$ and
\begin{equation}\label{mproj}
  \PQ^u_v=\delta^u_v - \frac{2k_{1}^{u} k_{1 v} + 2k_{2}^{u} k_{2 v}}{m^{2}}~.
\end{equation}
From \eqref{hq} it is easy to see that $\h_{uv}$ satisfies
\begin{equation}
  \h_{uv}k^v_\lambda = 0\ ,\qquad
  \h_{uv}h^{vw}\h_{wr} = \h_{ur}\ , \label{hid}
\end{equation}
where $h^{vw}$ is the inverse metric of the original quaternionic
manifold $\M_{\rm h}$, i.e.\ $h^{vw} h_{wu} = \delta_u^v$. We can then
use \eqref{mproj} to define the inverse metric on the quotient as
$\h^{uv} = \PQ^u_w h^{wv}$. The first equation in \eqref{hid} states
that the rank of $\h_{uv}$ is reduced by two relative to $h_{uv}$, which precisely
corresponds to the two Goldstone bosons that have been integrated
out. The second equation in \eqref{hid} tells us that the inverse
metric on the quotient $\h^{uv}$ actually coincides
with the inverse of the original metric $h^{vw}$.

Consistency with $\cN=1$ supersymmetry requires that $\h_{uv}$ is a
K\"ahler metric.  In order to show this we first need to find the
integrable complex structure on the K\"ahler manifold. It seems likely
that one of the three almost complex structures of the quaternionic
manifold descends to the complex structure on the quotient. Indeed, due to the $SU(2)$ gauge choice \eqref{P3},
$J^3$ plays a preferred role in that it points in the direction (in
$SU(2)$-space) normal to the plane spanned by $P_1^x, P_2^x$ and is
left invariant by the $U(1)$ rotation in that plane. One way to
calculate $J^3$ on the quotient is to employ the same method that we
just used for the metric and apply it to the two-form $K^3_{uv}$. This
is possible in an (auxiliary) two-dimensional $\sigma$-model of the
form\footnote{This Lagrangian has nothing to do with the theory
  considered so far and is only used to derive the form of the complex
  structure -- or rather its associated fundamental two-form -- on the
  quotient.  We thank E.\ Zaslow for suggesting this procedure.}
\begin{equation}\label{Laux} {\cal L}_{K^3} =
  K^3_{uv}  D_{\alpha} q^u D_{\beta} q^v\epsilon^{\alpha\beta}\ ,\quad
  \alpha, \beta = 1,2\ ,
\end{equation}
where the covariant derivatives are again given by \eqref{d2}.  As
above, we derive the algebraic equation of motion for
$A_\alpha^{\lambda}$ and insert it back into \eqref{Laux} to arrive at
\begin{equation} {\cal L}_{K^3} = \K_{uv} \epsilon^{\alpha\beta}
  \partial_{\alpha} q^u
  \partial_{\beta} q^v\ ,
\end{equation}
where
\begin{equation}\begin{aligned}\label{Khatdef}
\K_{uv} = K^3_{uv} - \frac{2k_{2 u} k_{1 v}-2k_{1 u} k_{2 v}}{m^2} =   \PQ^w_u K^3_{wv} \ .
\end{aligned}
\end{equation}
Here we have used the relations \eqref{Jk} to conclude that
$k^u_\lambda K^3_{uv} k^v_\rho = m^2 \epsilon_{\lambda\rho}$, where
$\epsilon_{21}=1$.
We find that the rank of $\K_{uv}$ is reduced by two due to
$k^u_{\lambda} \K_{uv} = 0$, analogous to the result for the metric
$h_{\mu\nu}$.

For two commuting isometries $k_1$ and $k_2$ we have the identity
\cite{Andrianopoli:1996cm}
\begin{equation}\label{equivariance_cond}
  2 k_1^u k_2^v K^x_{uv} + \epsilon^{xyz} P^y_1 P^z_2 = 0  \ ,
\end{equation}
which, together with \eqref {jhol}, allows us to simplify the
expression for the mass:
\begin{equation}\label{mass_prepot}
  m^2= P^1_1 P^2_2 - P^1_2 P^2_1 \ .
\end{equation}
On the other hand, from the definition of the prepotentials
\eqref{Pdef} we find
\begin{equation} \label{Killing_oneform}
  \begin{aligned}
    & k_{2v} ~=~  k^u_1 \,K_{uv}^3 ~ = ~  \omega^2_v P^1_1 - \omega^1_v P^2_1 \ , \\
    & k_{1v} ~=~ k^u_2 \,K_{uv}^3 ~ = ~ \omega^1_v P^2_2 - \omega^2_v
    P^1_2 \ ,
  \end{aligned}
\end{equation}
where we have used \eqref{P3} and \eqref{Jk}. Inserting \eqref{mass_prepot}  and
\eqref{Killing_oneform} into \eqref{Khatdef} we arrive at
\begin{equation}\label{Kform}
  \K_{uv} = \partial_u \omega_v^3 - \partial_v \omega_u^3\ .
\end{equation}
Thus, on ${\Mq}_{\rm h}$ there exists a fundamental two-form $\K$
which is indeed closed:
\begin{equation}\label{juhu}
  \diff\K = 0 \ .
\end{equation}
Furthermore, we find that $\J$ defined via $\K_{uv} = \h_{uw} \J^w_v$
is the projected complex structure $J^3$, i.e.
\begin{equation}
  \J^u_v = \PQ^u_w J^{3w}_{v}~.
\end{equation}
As $\PQ$ commutes with $J^3$, due to \eqref{jhol}, $\J$ is the
associated complex structure, i.e.\ it satisfies $\J^u_v \J^v_w = - \PQ^u_w$, which on the quotient reads $\J^2 =-{\bf  1}$. This, together with \eqref{juhu}, implies that the Nijenhuis-tensor $N(\J)$ vanishes. This completes the proof that ${\Mq}_{\rm h}$ is a K\"ahler manifold, with K\"ahler form $\K$ and complex structure $\J$.

In order to display the K\"ahler potential on ${\Mq}_{\rm h}$ let us explicitly introduce
complex coordinates. Since $\J$ is an honest complex structure, we can
group the $4n_{\rm h}-2$ coordinates $q^u$ into two sets of
coordinates $q^{2a-1}$ and $q^{2b}, a,b=1,\ldots,2n_{\rm h}-1$ such
that $\J$ is constant and `block-diagonal' in this basis, taking the
form
\begin{equation}
  \J_u^v =  \left(\begin{array}{ccccc} 0 & -1 && \\ 1 &0&&& \\ && \ddots && \\ &&&0 & -1  \\ &&&1 &0 \end{array}\right)\ .
\end{equation}
We can then define complex coordinates
\begin{equation}\label{zdef}
  z^a := q^{2a-1} +\iu q^{2a}\ , \quad \bar z^{\bar a} := q^{2a-1} - \iu q^{2a}\ ,
\end{equation}
and the associated derivatives
\begin{equation}\label{dzdef}
  \partial_{a} =
  \tfrac12\big(\partial_{q^{2a-1}} - \iu \partial_{q^{2a}}\big)\ ,\qquad
  \bar\partial_{\bar a} =
  \tfrac12\big(\partial_{q^{2a-1}} + \iu \partial_{q^{2a}}\big)\ .
\end{equation}
From $\J^w_u \J^t_v \K_{wt} = \K_{uv}$ we see that, in terms of
complex coordinates, the two-form $\K_{uv}$ given in \eqref{Kform} has
no $(2,0)$ and $(0,2)$ parts. In other words, $\K_{ab}=
\partial_a\omega_b^3-\partial_b\omega_a^3=0$ and $\K_{\bar a\bar b}=
\bar \partial_{\bar a} \bar \omega_{\bar b}^3-\bar \partial_{\bar b}
\bar \omega_{\bar a}^3=0$ . This in turn implies
\begin{equation}\label{Kahlerconnection}
  \omega_a^3 = \tfrac\iu2 \partial_a \K\ ,\qquad
  \bar\omega_{\bar a}^3 = -\tfrac\iu2 \bar\partial_{\bar a} \K\ ,
\end{equation}
where $\K$ is the (real) ${\cN=1}$ K\"ahler potential.\footnote{Note that
  one could add a further term in \eqref{Kahlerconnection} that does
  not contribute in \eqref{Kform} and corresponds to a K\"ahler
  transformation.}  Inserting these expressions into
\eqref{Kform} one obtains the K\"ahler-form
\begin{equation}\label{Kdef}
  \K_{a\bar b} =  \partial_a\bar\omega_{\bar b}^3-\bar\partial_{\bar b}\omega_a^3
  = -\iu\partial_a\bar\partial_b \K\ .
\end{equation}

So far, we have only integrated out the two vector bosons of the
massive gravitino multiplet including their Goldstone degrees of
freedom.  As we have just shown, the removal of the two Goldstone bosons
amounts to taking the quotient of the original quaternionic-K\"ahler
manifold ${\M}_{\rm h}$ with respect to the two gauged isometries
$k_{1,2}$. This quotient ${\Mq}_{\rm h}= {\M}_{\rm h}/<k_1,k_2>$ has
co-dimension two and is indeed a K\"ahler manifold, consistent with the
unbroken $\cN=1$ supersymmetry.  However, additional scalars from both vector- and/or hypermultiplets can acquire a mass of ${\cal O}(\mino)$ due to the partial supersymmetry breaking. Integrating out these scalar fields results in a submanifold ${\Mq}_{\rm v}$ of the original $\cN=2$ special-K\"ahler manifold ${\M}_{\rm v}$ and a submanifold of ${\Mq}_{\rm h}$. Thus, the final $\cN=1$ field space is the K\"ahler manifold
\begin{equation}\label{NoneMod}
 \M^{\cN=1} = {\Mq}_{\rm  v} \times {\Mq}_{\rm h}
\end{equation}
with K\"ahler potential
\begin{equation}\label{Kone}
  K^{\cN=1} = \hat K^{\rm v} +\K \ .
\end{equation}

Before we continue let us note that
the quotient construction presented in this section can also be
understood in terms of the corresponding
superconformal supergravity \cite{deWit:1984px,Kallosh:2000ve} or, equivalently,
in terms of the  hyper-K\"ahler cone construction
\cite{Hitchin:1986ea,Swann:1991,Gibbons:1998xa,deWit:1999fp}. In the $\cN=2$ superconformal theory,
the scalar field space of the hypermultiplets is given by a
$(4 n_{\rm  h}+4)$-dimensional hyper-K\"ahler cone ${\M}_{\rm HKC}$ over a $(4 n_{\rm  h}+3)$-dimensional tri-Sasakian manifold, which itself is an $S^3$-fibration over the quaternionic base ${\M}_{\rm h}$. Thus, ${\M}_{\rm h}$ can be viewed as the quotient ${\M}_{\rm h} = {\M}_{\rm HKC}/(SU(2)_R\times\mathbb{R}_+)$, where dilatations and $SU(2)_R$ act on the cone and fibre directions, respectively. ${\M}_{\rm HKC}$ is hyper-K\"ahler and thus has
three integrable complex structures which descend to the three almost
complex structures $J^x$ on ${\M}_{\rm h}$.

In the superconformal framework partial supersymmetry breaking would correspond to taking a K\"ahler quotient of ${\M}_{\rm HKC}$ with respect to the holomorphic Killing vector $k_1 + \iu k_2$ to produce an $\cN=1$ superconformal theory. On this K\"ahler quotient
only one of the three complex structures should be well-defined and thus
$SU(2)_R$ is broken to $U(1)_R$. In other words, the fibre $S^3$ is projected onto an $S^1$ on which the $\cN=1$ $U(1)_R$ acts, while the cone direction $\mathbb{R}_+$ is not effected. Therefore, when $\cN=2$ to $\cN=1$ supersymmetry breaking occurs in superconformal supergravity, a minimum of four scalars should be removed from the spectrum - two are eaten by the gauge bosons in the massive
gravitino multiplet and two are eaten by the massive $SU(2)_R$ gauge bosons. The structure of the $\cN=1$ superconformal theory then implies that we have a rigid K\"ahler manifold of dimension
$4 n_{\rm h}$ which is an $\mathbb{R}_+$ cone over a
$4 n_{\rm  h}-1$ dimensional Sasakian manifold, which itself is a $S^1$-fibration over a $4 n_{\rm  h}-2$ K\"ahler base ${\Mq}_{\rm h}$ \cite{Gibbons:1998xa,Kallosh:2000ve}. Fixing the superconformal symmetry corresponds to taking the standard K\"ahler quotient, i.e.\ gauge fixing the dilatation ($\mathbb{R}_+$) and the $U(1)_R$ , to leave ${\Mq}_{\rm h}$ as the $\cN=1$ scalar field space of the effective theory, which is K\"ahler by construction.

We will not study the superconformal version of partial supersymmetry breaking in any further detail here. However, in Section \ref{section:SQC} we shall see that a knowledge of the hyper-K\"ahler cone construction proves useful in determining the K\"ahler potential and the holomorphic coordinates on ${\Mq}_{\rm h}$.

\subsection{The gauge couplings}
\label{section:Nonef}

Let us now check the holomorphicity of the gauge couplings.  In
section~\ref{section:NoneK} we integrated out the two heavy gauge
bosons in the low-energy limit by neglecting their kinetic terms and
using their algebraic equations of motion. In order to compute the
gauge couplings of the light gauge fields that descend to
the $\cN=1$ theory we have to explicitly project out the heavy gauge
bosons in the coupled kinetic terms in \eqref{sigmaint}. From
\eqref{Acomb} we see that the projection is determined by the
embedding tensor solutions given in \eqref{solution_embedding_tensor2} and \eqref{solution_embedding_tensor_AdS2}. In other
words, we should impose the projection
\begin{equation}\label{Fcond}
  \Theta^{\lambda I} G\,_I^\pm +  \Theta^{\lambda}_I\, F^{I \pm} = 0\ ,\qquad \lambda=1,2
\end{equation}
and then compute the gauge couplings of the remaining gauge fields.
Taking complex combinations and inserting the embedding tensor solutions
\eqref{solution_embedding_tensor2} yields\footnote{We only discuss the
  Minkowski case here. The AdS case is completely equivalent, in that
  \eqref{solution_embedding_tensor_AdS2} only leads to a different
  prefactor (i.e.\ not $C^I$) but the conclusion remains the same.}
\begin{equation}\label{Fint}
  C^I (\cF_{IJ}(\hat{t}) -  \mathcal{N}_{IJ}(\hat{t})) F^{J +} ~=~ 0 ~ =~ \bar{C}^I
  (\bar{\cF}_{IJ}(\hat{t}) -  \mathcal{N}_{IJ}(\hat{t})) F^{J +}\ ,
\end{equation}
and a similar set of equations for $F^{J -}$. Note that
$\cF_{IJ}$ and $\mathcal{N}_{IJ}$ are evaluated in the $\cN=1$ background, which means that
scalar fields not obeying \eqref{deformV} are fixed at their
background values. The scalars $\hat{t}$ of the $\cN=1$ theory, which do obey \eqref{deformV}, can vary
arbitrarily.

Using the definition of $\mathcal{N}_{IJ}$ \eqref{Ndef} we find that \eqref{Fint} implies
\begin{equation}\label{Fproj}
  X^I \Im\left(\cF_{IJ}(\hat{t}) \right) F^{J +} = 0  ,
\end{equation}
where we have dropped a non-vanishing prefactor.
This condition projects out one linear combination of the $F^{I}$
that is heavy. For the following analysis it will be useful to define
the related projection operator
\begin{equation}\label{Pidef}
  \bar \Pi^I_J \equiv \delta^I_J + 2 \e^{K^{\rm v}} \bar X^I X^K \Im(\cF)_{KJ} \ ,
\end{equation}
such that $(1-\bar\Pi)$ projects onto the heavy gauge boson while
$\bar\Pi$ projects onto the orthogonal gauge bosons. Note that in \eqref{Pidef} (and from now on) we have dropped the explicit $\hat{t}$-dependence for convenience.

Before we identify the second heavy gauge boson let us check which
physical field is projected out by \eqref{Fproj}. Looking at the full
$\cN=2$ gravitino variation \cite{Andrianopoli:1996cm}, we see that it
contains the `dressed' graviphoton term
\begin{equation}
  \tilde T_{\mu\nu}^{+} = 2 \iu \bar X^{I} \Im  \mathcal{N}_{IJ} F^{J + }_{\mu\nu} + \ldots~.
\end{equation}
It is straightforward to check that the projection $\bar X^{I}
\Im\mathcal{N}_{IJ}$ appearing here coincides with
\eqref{Fproj} \cite{Ceresole:1995ca}. Therefore, \eqref{Fproj} can be understood as
projecting out the graviphoton.

The second projection condition implied by \eqref{Fint} reads
\begin{equation} \label{FprojC}
C^{(P)\,I} \Im(\cF)_{JK} F^{K+} = 0 \ ,
\end{equation}
where we have defined $C^{(P)\,I}= \Pi^I_JC^J$.  Expressing this in terms of
the projection operator
\begin{equation}\label{Gammadef}
  \bar \Gamma_{J}^I \equiv \delta^I_J - \frac{\bar  C^{(P)\,I}   C^{(P)\,K}  \Im(\cF)_{KJ}}{ C^{(P)\,M}  \Im(\cF)_{MN} \bar  C^{(P)\,N}} \ ,
\end{equation}
we see that $(1-\bar\Gamma)$ projects onto the second heavy gauge
boson while $\bar\Gamma$ projects to the orthogonal gauge bosons.
With the help of the two projection operators, which one can show commute, we are now in the
position to define the light vector fields which remain in the $\cN=1$
theory by
\begin{equation} \label{FN=1} F^{\hat I +} \equiv F^{I +}\Big|_{\cN=1}
  = \bar \Pi^I_J \bar \Gamma^{J}_K F^{K +} \ ,
\end{equation}
where \ $\hat I = 1,\dots n'_v = (n_{\mathrm v}-1)$, i.e.\ we have
projected out two of the $\cN=2$ vectors. In Appendix \ref{section:massive_multi} we further check that the masses of the two heavy gauge bosons obey the $\cN=1$ relations with the gravitino mass.

Let us now return to our original task and compute the gauge coupling
functions of the $\cN=1$ action. This can be done by imposing the two
projections \eqref{Fproj} and \eqref{FprojC} on the gauge kinetic term
$\mathcal{N}_{IJ}\,F^{I +}_{\mu\nu}F^{\mu\nu\, J+}$ of
\eqref{sigmaint}. In other words, we should compute
$\mathcal{N}_{\hat I\hat J}\,F^{\hat I +}_{\mu\nu}F^{\mu\nu\, \hat
  J+}$ with $F^{\hat I +}$ given by \eqref{FN=1}. Inserting the
definition of $\mathcal{N}_{I J}$ \eqref{Ndef} we find that the
$\cN=1$ gauge coupling functions appearing in \eqref{N=1Lagrangian}
are given by
\begin{equation}\label{fN=1}
  \bar f_{\hat I\hat J}(\hat{t}) = -\mathrm{i} \bar \cF_{\hat I\hat J}\  \ ,
\end{equation}
where the second term in \eqref{Ndef} drops out due to the
identity
\begin{equation}\label{Fproj2}
  X^I \Im(\cF)_{I\hat J} F^{\hat J +} = 0 \ .
\end{equation}
It is straightforward to see that \eqref{Fproj2} holds by inserting \eqref{FN=1} and using $e^{-K_{\rm v}}
= -2 \bar X^I {\rm Im}(\cF)_{IJ} X^J$.

As promised, we see that the gauge couplings are manifestly
holomorphic. Furthermore, $f_{\hat I\hat J}(t)$ can only depend on the
scalar fields that descend from $\cN=2$ vector multiplets, but not on
those descending from hypermultiplets. In fact, this is analogous to
the situation in $\cN=2\to \cN=1$ truncations, where the graviphoton
also has to be projected out and, as a consequence, the gauge couplings
are holomorphic and only depend on the scalars of the vector
multiplets \cite{Andrianopoli:2001zh,Andrianopoli:2001gm}.

\subsection{The superpotential}
\label{section:NoneW}

Our next task is to determine the $\cN=1$ superpotential ${\cal W}$.
This is most easily done by comparing the supersymmetry transformation of the
$\cN=1$ gravitino $\Psi_{\mu\, 1}$ \eqref{susytrans2} with
the conventional $\cN=1$ transformation given, for example, in
\cite{Wess:1992cp}. (An analogous computation
for $\cN=1$ truncations of $\cN=2$ theories can be found in
\cite{Grana:2005ny,Andrianopoli:2001zh,Andrianopoli:2001gm}).  Focusing
on the scalar contribution one has
\begin{equation}\label{Nonegravitino}
  \delta_\epsilon \Psi_{\mu\,1} = D_\mu \epsilon - S_{11}
  \gamma_\mu \bar \epsilon  + \ldots \,\ =\ D_\mu \epsilon - \tfrac12 \e^{\tfrac12K^{\cN=1}} {\cal W}
  \gamma_\mu \bar \epsilon  + \ldots
\end{equation}
where we have already inserted our choice $\epsilon_1
= {\epsilon\choose 0}$ and the right-hand side is the $\cN=1$ gravitino variation expressed in terms of the superpotential ${\cal W}$.

Using the definition of the gravitino mass matrices \eqref{susytrans3}
we find that the $\cN=1$ superpotential is given by
\begin{equation}\label{Wone}
  {\cal W} =  2 \e^{-\tfrac12K^{\cN=1}} S_{11}\ =\ \e^{-\K/2} {V}^\Lambda
  \Theta_\Lambda^{~~\lambda} P_{\lambda}^- \ .
\end{equation}
In this expression we have to appropriately project out all scalars
with masses of ${\cal O}(\mino)$. In other words, ${\cal W}$ should be
expressed in terms of $\cN=2$ input couplings restricted to the light
$\cN=1$ modes. As we discussed at the end of section \ref{section:NoneK},
this projection preserves the K\"ahler and complex structure
of ${\M}_{\rm v} \times {\Mq}_{\rm h}$. Therefore, we should be able to
check the holomorphicity of ${\cal W}$ without knowing the precise
$\cN=1$ spectrum.

Before continuing,  let us
discuss the situation where the original $\cN=2$
supergravity is only gauged with respect to the two Killing vectors
$k_1,k_2$ that induce the partial breaking.
In this case the index $\lambda$
in \eqref{Wone} takes the values $\lambda=1,2$ and
all  fields in the $\cN=1$ effective theory are exactly massless, i.e.\ they are $\cN=1$ moduli. Their vacuum expectation
values are not fixed, or, in other words, they parametrise the entire
$\cN=1$ background. As a consequence the superpotential has to be proportional
to the cosmological constant.  This can be seen explicitly by inserting the gravitino mass matrix
\eqref{N=1conditions} into \eqref{Wone} which gives
\begin{equation}
|{\cal  W}|^2= 4 \e^{-K^{\cN=1}} |S_{11}|^2 = 4 \e^{-K^{\cN=1}} |\mu|^2 \ ,
\end{equation}
in agreement with the standard $\cN=1$ relation \cite{Wess:1992cp}.

If an additional $m$ Killing vectors are gauged, then their corresponding Killing prepotentials appear in \eqref{Wone}  and the index $\lambda$ runs over all $m+2$ values. For this case we will now show that ${\cal W}$ is holomorphic with respect to the $\cN=1$ complex structure determined in the previous section.

Inspecting the superpotential ${\cal W}$ given in \eqref{Wone} we see that the scalars of ${\M}_{\rm v}$ already appear holomorphically via $V^\Lambda$. Therefore, we are left to show that the anti-holomorphic derivative of ${\cal W}$ with respect to the scalars of ${\Mq}_{\rm h}$ vanishes, i.e.\
\begin{equation}\label{der_W}
  \bar\partial_{\bar a} {\cal W} = \e^{-\K/2} {V}^\Lambda
  \Theta_\Lambda^{~~\lambda} (\bar \partial_{\bar a} P_\lambda^- - \tfrac12 (\bar \partial_{\bar a} \K)P_\lambda^-)  = 0\ .
\end{equation}
Let us first note that using \eqref{Kahlerconnection} we can
express $\bar \partial_{\bar a} \K$ in terms of $\omega^3_{\bar a}$.
Furthermore, from the definition of Killing prepotentials \eqref{Pdef}
we see that
\begin{equation}
  -2 K^-_{uv} k^v_\lambda = \partial_u P^-_\lambda + \iu \omega^-_u P^3_\lambda - \iu \omega^3_u P^-_\lambda \ ,
\end{equation}
which implies
\begin{equation}\label{der_W2}
  \bar\partial_{\bar a} {\cal W} = - \e^{-\K/2} {V}^\Lambda
  \Theta_\Lambda^{~~\lambda} (2 K^-_{\bar a v} k^v_\lambda + \iu \bar \omega^-_{\bar a} P_\lambda^3) \ .
\end{equation}
From the quaternionic algebra \eqref{jrel} and $K^x_{uv} = h_{uw} (J^x)^w_v$ it is easy to see that $K^-$ is actually a $(2,0)$-form and thus only has  holomorphic indices. This immediately implies that the first term in the bracket vanishes. From \eqref{Killing_oneform} we can infer that both $\omega^1$ and $\omega^2$ live entirely in the space spanned by $k_{1v}$ and $k_{2v}$, which in fact is divided out. This implies that $\omega^-_{\bar a}$ is zero on ${\Mq}_{\rm h}$ and therefore the second term in \eqref{der_W2} also vanishes. Thus, the superpotential ${\cal W}$ is holomorphic, consistent with $\cN=1$ supersymmetry.


\subsection{The D-terms}
\label{section:NoneD}

Our final task is to explicitly compute the $\cN=1$ D-terms appearing in the effective potential \eqref{N=1pot}. This proceeds analogously to the calculation of the superpotential in Section \ref{section:NoneW}, but by comparing the $\cN=2$ and $\cN=1$ gaugino variations instead of the gravitino variations. Once again, this procedure is similar that used in $\cN=1$ truncations \cite{Andrianopoli:2001zh,Andrianopoli:2001gm,D'Auria:2005yg}, but here we shall more closely follow \cite{Cassani:2007pq}.

The $\cN=2$ gaugino variation is given by \cite{Andrianopoli:1996cm}
\begin{equation}\label{gauginivar}
  \delta_\epsilon \lambda^{i {\cal A}} = \gamma^\mu\partial_\mu t^i \epsilon^{\cal A} - \tilde G_{\mu\nu}^{i -} \gamma^{\mu\nu} \varepsilon^{\cal AB}\epsilon_{\cal B} + W^{i{\cal AB}}\epsilon_{\cal B}+\ldots~,
\end{equation}
where $W^{i {\cal AB}}$ was defined in \eqref{susytrans3} and $\tilde
G_{\mu\nu}^{i -} = -g^{i\bar j} \nabla_{\bar j}\bar X^{I}
\mathrm{Im}\mathcal{N}_{IJ} F^{J - }_{\mu\nu} + \ldots$ are the
`dressed' anti-self-dual field strengths, with the ellipses denoting higher-order fermionic contributions.

In order to identify the gaugini of the effective $\cN=1$ theory we evaluate \eqref{gauginivar} for our choice of the preserved supersymmetry parameter $\epsilon_1 = {\epsilon\choose 0}$ and obtain
\begin{eqnarray}\begin{aligned}
  \delta_\epsilon \lambda^{i 1} ~=&~ \gamma^\mu\partial_\mu t^i
\bar\epsilon +  W^{i{11}} \epsilon~ +\ldots~, \label{trunc-fermion} \\
  \delta_\epsilon \lambda^{i 2} ~=&~ - \tilde G_{\mu\nu}^{i -}
\gamma^{\mu\nu} \epsilon +  W^{i{ 21}}  \epsilon~ +\ldots~. \label{trunc-gaugino}
\end{aligned}
\end{eqnarray}
Comparing with the standard $\cN=1$ gaugino variation \cite{Wess:1992cp,Gates:1983nr}
\begin{equation}
  \delta_\epsilon \lambda^{\hat{I}} =  F_{\mu\nu}^{\hat{I} -}
\gamma^{\mu\nu} \epsilon + \mathrm{i} {\cal D}^{\hat I} \epsilon~ +\ldots~, \label{NoneGauginoVar}
\end{equation}
we conclude that the $\lambda^{i 2}$ are candidates for $\cN=1$ gaugini.  However, not all $\lambda^{i 2}$ descend to the effective $\cN=1$ theory as some of them are massive and have to be integrated out. The $\cN=1$ gaugini should be defined as those with the light $\cN=1$ gauge fields \eqref{FN=1} appearing in their supersymmetry variations. Using the projection operators $\Pi$ and $\Gamma$, given in \eqref{Pidef} and \eqref{Gammadef} respectively, and the definition \eqref{FN=1} we can restrict the gauge fields appearing in the $\cN=2$ gaugino variation \eqref{trunc-gaugino} to the light $\cN=1$ gauge fields. By comparing the resulting expression with the $\cN=1$ gaugino variation \eqref{NoneGauginoVar}, we can identify the $\cN=1$ gaugini as
\begin{equation}\label{NoneGaugino}
  \lambda^{\hat{I}} = -2e^{K^{\rm v}/2} \nabla_{i} X^{\hat I} \lambda^{i 2}   ~,
\end{equation}
where we have used the same projector \eqref{FN=1} to define
\begin{equation}\label{dxproj}
\nabla_{i}X^{\hat I} = \Pi^I_J  \Gamma^{J}_K \nabla_{i}X^K~.
\end{equation}
In order to reach the result \eqref{NoneGaugino}, we have first made use of the special-geometry
relation \cite{Andrianopoli:1996cm}
\begin{equation}\label{special}
  \nabla_{i} X^{\hat I}  g^{i\bar \jmath}\, \nabla_{\bar \jmath}\bar {X}^{\hat J} = -\tfrac{1}{2} e^{-K^{\rm v}}(\mathrm{Im} \cN)^{-1 ~\hat I \hat J} - X^{\hat I}  \bar {X}^{\hat J}~,
\end{equation}
which is derived from the standard identity restricted to the light $\cN=1$ fields using the projection operators $\Pi$ and $\Gamma$ and \eqref{dxproj}. We can then simplify \eqref{special} by making use of the fact that the projector $\Pi$ given in \eqref{Pidef} is defined such that the following property holds\footnote{Note that \eqref{Xcond} does not fix any scalars, as the projection operators $\Pi^I_J$ and $\Gamma^{J}_K$ are field-dependent quantities which vary over the $\cN=1$ moduli space. This should be compared to $\cN=2 \rightarrow \cN=1$ supergravity truncations\cite{Andrianopoli:2001zh,Andrianopoli:2001gm}, where the equivalent projection operators are constant and, therefore, some scalars are fixed by the condition $\Pi^I_J  X^J =0$.}
\begin{equation}\label{Xcond}
  X^{\hat I}= \Pi^I_J  \Gamma^{J}_K X^K =0~,
\end{equation}

We can now take the $\cN=1$ supersymmetry variation of \eqref{NoneGaugino}
(to lowest fermionic order), use \eqref{trunc-gaugino}, insert the
definition of $W^{i{ 21}}$ \eqref{susytrans3}, and compare the result
with the standard $\cN=1$ expression \eqref{NoneGauginoVar} to read
off the D-term:
\begin{eqnarray}\begin{aligned}
  {\cal D}^{\hat I} ~=&~ 2\mathrm{i}e^{K^{\rm v}/2} \nabla_{i} X^{\hat I} W^{i{ 21}} \nonumber \\
  ~=&~-2 e^{K^{\rm v}} \nabla_{i} X^{\hat I} g^{i\bar \jmath}\,
  \nabla_{\bar \jmath}\bar {X}^{\hat J}\left( \Theta_{\hat
      J}^{~~\lambda} - {\cal N}_{\hat J \hat K} \Theta^{\hat
      K\lambda}\right) P_{\lambda}^3 ~, \label{Dterm}
\end{aligned}
\end{eqnarray}
where we have used $\nabla_{i} F_{\hat J} = {\cal F}_{\hat J \hat K}\nabla_{i} X^{\hat K}$ in the second line. In order to see that this expression agrees with the standard $\cN=1$ D-term \eqref{NoneDterm}, we again make use of \eqref{special} and \eqref{Xcond} to see that it can be written as
\begin{equation}\label{D_terms}
  {\cal D}^{\hat I} = - (\mathrm{Re} f)^{-1 ~\hat I \hat J} \left( \Theta_{\hat J}^{~~\lambda} - \iu \bar{f}_{\hat J \hat K} \Theta^{\hat K\lambda}\right) P_{\lambda}^3~.
\end{equation}
Therefore we can identify the $\cN=1$ Killing prepotential as follows
\begin{equation}\label{N=1prepotential} {\cal P}_{\hat J} =
  \tfrac{1}{2} \left( \Theta_{\hat J}^{~~\lambda} - \iu \bar{f}_{\hat J \hat K}
    \Theta^{\hat K\lambda}\right) P_{\lambda}^3~.
\end{equation}
If we now consider gaugings with respect to just the Killing vectors $k_1$ and $k_2$ responsible for partial supersymmetry breaking, we see that the D-term vanishes by our $\cN=1$ supersymmetry condition \eqref{P3}, as expected for a supersymmetric vacuum.

Note that both the D-terms \eqref{D_terms} and the Killing prepotentials \eqref{N=1prepotential} are complex,
in agreement with the analogous results from $\cN=1$ truncations \cite{D'Auria:2005yg,Cassani:2007pq}. The reason is that these quantities appear in the supersymmetry variations of the gaugini in \eqref{trunc-gaugino} which are paired with the (complexified) anti-self-dual field strengths $\tilde G_{\mu\nu}^{i -}$. Therefore, \eqref{D_terms} describes a complex linear combination of the electric and the magnetic D-terms. More precisely, from \eqref{N=1prepotential} we see that the electric and magnetic Killing prepotentials of the $\cN=1$ theory are given by $\frac{1}{2} \Theta_{\hat J}^{~~\lambda} P_{\lambda}^3$ and $\frac{1}{2} \Theta^{\hat K\lambda} P_{\lambda}^3$.\footnote{We thank D.\ Cassani and G.\ Dall'Agata for useful discussions on this point.}

Before we close this section let us
note that one can also check that the supersymmetry
transformation of the $\cN=1$ fermions in chiral multiplets that descend
from the $\cN=2$ gaugini $\lambda^{i1}$ (cf.~\eqref{trunc-fermion})
correctly reproduces the F-terms. Furthermore,
one might expect that it is necessary to take field redefinitions of
the gaugini and the hyperini with respect to the Goldstino, such that
we can rewrite the fermionic Lagrangian in terms of physical fermions,
i.e.\ fermions that cannot be gauged away by further field
redefinitions of the massive gravitino $\Psi_{\mu 2}$ \cite{Gunara:2003td}.
However, it is
straightforward to check that any such field redefinitions are
projected out when one identifies the $\cN=1$ fields as in
\eqref{NoneGaugino}. In other words, the $\cN=1$ fermionic field space
is defined by quotienting the $\cN=2$ counterpart by the Goldstino
direction.

This completes our analysis of the low-energy effective theory in the
partial supersymmetry breaking vacua of $\cN=2$ gauged supergravity
with electric and magnetic charges. We have proven that this theory
enjoys $\cN=1$ supersymmetry, as is required for the consistency of
the partial supersymmetry breaking mechanism. We shall now focus on the class of special quaternionic-K\"ahler manifolds.

\section{Special quaternionic-K\"ahler manifolds}
\label{section:SQC}

In this section we will provide an explicit example of the results of Section \ref{section:None} by deriving the $\cN=1$ effective action for the class of supergravities that arise at string tree-level in type II compactifications.
In this case the $4n_{\rm h}$-dimensional quaternionic-K\"ahler manifold $\M_{\rm h}$ takes a special form, in that its metric is entirely determined in terms of the holomorphic prepotential of a
$(2n_{\rm h}-2)$--dimensional special-K\"ahler submanifold $\M_{\rm sk}$.
Such a manifold $\M_{\rm h}$ is called special quaternionic-K\"ahler and the construction of its metric is known as the c-map \cite{Cecotti:1988qn,Ferrara:1989ik}. In \cite{Louis:2009xd} we showed that
$\cN=1$ vacua generically exist for
this subclass of quaternionic K\"ahler manifolds.
In the following we will determine the K\"ahler potential, the superpotential and the D-terms of the corresponding effective action.

Let us denote the complex coordinates of $\M_{\rm  sk}$ by
$z^a, a=1,\ldots,\nh-1,$ its K\"ahler potential by $K^{\rm h}(z,\bar z)$ and the holomorphic prepotential by ${\cal G}(z)$. The remaining scalars in the
hypermultiplets are the dilaton $\phi$, the axion $\ax$ and  $2\nh$ real Ramond-Ramond scalars  $\xi^A, \tilde\xi_A, A=1,\ldots,\nh$.
Together they define a $G$-bundle over $\M_{\rm  sk}$, where $G$ is the semidirect product of a $(2\nh+1)$-dimensional Heisenberg group with $\mathbb{R}$. The Killing vectors corresponding to the action of $G$ can be used to construct $\cN=1$ solutions \cite{Louis:2009xd}.

In \cite{Ferrara:1989ik} it was observed that there is a specific parametrisation of the quaternionic vielbein $\mathcal U^{\mathcal A\alpha}_u$, defined in \eqref{Udef}, which reads (our notation follows \cite{Cassani:2007pq,Louis:2009xd})
\begin{equation} \label{quat_vielbein}
\mathcal U^{\mathcal A\alpha}= \mathcal U^{\mathcal A\alpha}_u \diff q^u =\tfrac{1}{\sqrt{2}}
\left(\begin{aligned}
 \bar{u} && \bar{e} && -v && -E \\
\bar{v} && \bar{E} && u && e
\end{aligned}\right) \ ,
 \end{equation}
where the one-forms are defined by
\begin{equation} \label{one-forms_quat}
 \begin{aligned}
  u ~= &~ \iu \e^{K^{\rm h}/2+\phi}Z^A(\diff \tilde\xi_A - \mathcal M_{AB} \diff \xi^B) \ , \\
  v ~= &~ \tfrac{1}{2} \e^{2\phi}\big[ \diff \e^{-2\phi}-\iu (\diff \ax +\tilde\xi_A \diff \xi^A-\xi^A \diff \tilde \xi_A  ) \big] \ , \\
  E^{\,\underline{b}} ~= &~ -\tfrac{\iu}{2} \e^{\phi-K^{\rm h}/2} {\Proj}_A^{\phantom{A}\underline{b}} (\Im  \mathcal G)^{-1\,AB}(\diff \tilde\xi_B - \mathcal M_{BC} \diff \xi^C) \ , \\
  e^{\,\underline{b}} ~= &~ {\Proj}_A^{\phantom{A}\underline{b}} \diff Z^A \ . \\
 \end{aligned}
\end{equation}
Here $Z^A$ are the homogeneous coordinates of $\M_{\rm  sk}$, ${\Proj}_A^{\phantom{A}\underline{b}} =(-e_a^{\phantom{a}\underline{b}}Z^a, e_a^{\phantom{a}\underline{b}})$ is defined using the vielbein $e_a^{\phantom{a}\underline{b}}$ on $\M_{\rm  sk}$ and
$\mathcal M_{AB}$ is computed from the prepotential ${\cal G}$ exactly as ${\cal N}_{IJ}$ is determined by ${\cal F}$ in \eqref{Ndef}.
The metric $h_{uv}$ on $\M_{\rm h}$ is
\begin{equation}
h = \left[ v \otimes \bar v + u \otimes \bar u + E \otimes \bar E  + e \otimes \bar e \right]_{\rm sym} \ .
\end{equation}
Given the explicit form of the vielbein \eqref{quat_vielbein} the $SU(2)$ connections $\omega^x$ reads \cite{Ferrara:1989ik}
\begin{equation}\label{quat_connection}
\begin{aligned}
\omega^1 ~= &~  \iu (\bar u- u)  \ , \qquad
\omega^2 ~= ~ u + \bar u \ , \\
\omega^3 ~= &~ \tfrac{\iu}{2} (v-\bar v) - \iu \e^{K^{\rm h}} \left(Z^A (\Im  \mathcal G)_{AB} \diff \bar Z^B - \bar Z^A (\Im  \mathcal G)_{AB} \diff Z^B \right) \ .
 \end{aligned}
\end{equation}

As already anticipated, the metric of $\M_{\rm h}$ has
$(2n_{\rm h}+2)$ isometries generated by the Killing vectors
\begin{equation}\label{Killing}
\begin{aligned}
\k_{\phi}   ~= &~ \tfrac{1}{2} \frac{\partial}{\partial \phi} -  \ax \frac{\partial}{\partial \ax} - \tfrac{1}{2} \xi^A \frac{\partial}{\partial \xi^A} - \tfrac{1}{2} \tilde \xi_A \frac{\partial}{\partial \tilde \xi_A} \ , \\
 k_{\tilde \phi}   ~= &~ - 2 \frac{\partial}{\partial \ax} \ , \\
 k_A  ~= &~ \frac{\partial}{\partial \xi^A} + \tilde \xi_A \frac{\partial}{\partial \ax} \ , \\
\tilde k^A  ~= &~ \frac{\partial}{\partial \tilde \xi_A} - \xi^A \frac{\partial}{\partial \ax} \ .
\end{aligned}
\end{equation}
The corresponding Killing prepotentials $P^x_\lambda$, defined in \eqref{Pdef}, take the following simple form
\cite{Michelson:1996pn,Cassani:2007pq}
\begin{equation} \label{prepotential_no_compensator}
 P^x_\lambda = \omega^x_u k_\lambda^u \ .
\end{equation}

After these preliminaries, we can explicitly compute the couplings of the $\cN=1$ effective action.
However, it will be necessary to discuss Minkowski and AdS backgrounds separately.
Let us start with the Minkowski case.

\subsection{Minkowski vacua} \label{section:Min_special}
In \cite{Louis:2009xd} we showed that the two Killing vectors needed
for partial supersymmetry breaking are given by
\begin{equation} \label{Killingv} \begin{aligned}
    k_1 =& \ \Im\big(D^A (k_A + {\cal G}_{AB} \tilde k^B )\big) +\Im\big(D^A ({\tilde \xi}_A - {\cal G}_{AB}  \xi^B )\big) k_{\tilde \phi}  \ , \\
    k_2=&\ \Re\big(D^A (k_A + {\cal G}_{AB} \tilde k^B
    )\big)+\Re\big(D^A ({\tilde \xi}_A - {\cal G}_{AB} \xi^B )\big)
    k_{\tilde \phi} \ ,
\end{aligned}\end{equation}
where $k_A, \tilde k^B$ and $k_{\tilde \phi}$ are defined in \eqref{Killing} and $D^A$ is a complex vector obeying
\begin{equation}
 \bar D^A (\Im {\cal G})_{AB} D^B = 0 \ .
\end{equation}
Furthermore, the prefactors in \eqref{Killingv} have to be constant in order for $k_1$ and $k_2$ to be Killing vectors, i.e.\
\begin{equation} \label{constant_Min}
 D^A = \textrm{const.} \ , \qquad
 D^A {\cal G}_{AB} = \textrm{const.} \ , \qquad
 D^A ({\tilde \xi}_A - {\cal G}_{AB} \xi^B ) = \textrm{const.} \ .
\end{equation}
The scalars that obey \eqref{constant_Min} define the $\cN=1$ locus, while those violating \eqref{constant_Min} have a mass of ${\cal O}(\mino)$. In the $\cN=1$ effective action all such massive fields are integrated out, which corresponds to setting the variation of the prefactors in \eqref{Killingv} to zero.
From the third equation in \eqref{constant_Min} we see that only two coordinates in the fibre are stabilised.
For the base coordinates the second equation in \eqref{constant_Min} implies
\begin{equation}\label{deformSQ}
{\cal G}_{ABC} D^B \delta Z^C = 0 \ .
\end{equation}
Analogously to \eqref{deformV}, for generic ${\cal G}$ this gives $n_{\rm h}-1$ complex conditions and thus stabilises all coordinates of $\M_\textrm{sk}$.

A special case occurs when the prepotential ${\cal G}$ is quadratic, which corresponds to
\begin{equation}
\M_\textrm{sk}=\frac{SU(1,n_{\rm h}-1)}{SU(n_{\rm h}-1)} \ , \qquad \M_\textrm{h}=\frac{U(n_{\rm h},2)}{U(n_{\rm h})\times U(2)} \ .
\end{equation}
Then \eqref{deformSQ} is trivially satisfied and all base coordinates together with $2n_{\rm h}-2$ fibre coordinates descend to a total of $4n_{\rm h}-4$ light scalar fields in the $\cN=1$ theory. In contrast, for a generic cubic prepotential the condition \eqref{deformSQ} stabilises all base coordinates, leaving an $\cN=1$ scalar field space of dimension $2n_{\rm h}-2$.\footnote{It is known that $\M_\textrm{sk}$ admits shift isometries for the imaginary parts of the $z^a$, therefore one might expect the $z^a$ to also be massless. However, these isometries generically induce symplectic transformations on the vector of fibre coordinates $\xi_{\tilde \lambda} = (\tilde \xi_A,\xi^A)$, see \cite{deWit:1992wf}. If $k_1$ and $k_2$ transform non-trivially under these symplectic transformations, the corresponding symmetries are broken by the gaugings and the related scalars gets a mass, indicated by the condition \eqref{deformSQ}. This is analogous to the discussion of isometries on $\M_{\rm v}$, see Section \ref{section:scales}.}

Let us now determine the couplings of the $\cN=1$ effective theory.
In order to apply the procedure developed in the previous section we should first check that the $SU(2)$ gauge choice \eqref{P3} holds. By inserting \eqref{Killingv} into \eqref{prepotential_no_compensator} we find
\begin{equation}\begin{aligned}
 P^3_1  ~= &~ \e^{2\phi} \Im D^A\Big( (\tilde \xi_A - \hat {\tilde \xi}_A) - {\cal G}_{AB}(\hat z) (\xi^B - \hat \xi^B) \Big) \ , \\
 P^3_2  ~= &~ \e^{2\phi} \Re D^A\Big( (\tilde \xi_A - \hat {\tilde \xi}_A) - {\cal G}_{AB}(\hat z) (\xi^B - \hat \xi^B) \Big) \ ,
\end{aligned}\end{equation}
where the coordinates $(\hat {\tilde \xi}_A,\hat \xi^A)$ and $\hat z$ parametrise the $\cN=1$ locus, while $(\tilde \xi_A,\xi^A)$ also include the massive scalars.
We see that in the $\cN=1$ locus $P^3_{1,2} = 0$ indeed holds but
$\diff P^3_{1,2} = 0$ is not fulfilled. More precisely, the one-forms
\begin{equation}\label{dP3_Mink}\begin{aligned}
\diff P^3_{1}  ~= &~ \e^{2\phi}\Im \Big( D^A (\diff \tilde \xi_A - {\cal G}_{AB}(z) \diff \xi^B ) \Big) \ , \\
\diff P^3_{2}  ~= &~ \e^{2\phi}\Re \Big( D^A (\diff \tilde \xi_A - {\cal G}_{AB}(z) \diff \xi^B ) \Big)
\end{aligned}\end{equation}
point in the direction of the massive scalars in the fibre.
Integrating out these scalars automatically sets $\diff P^3_{1,2} = 0$ and we recover the \eqref{P3}.

Let us now compute the K\"ahler two-form and the K\"ahler potential.
Using \eqref{Kform}, the $SU(2)$ connections \eqref{quat_connection} and the results for the exterior derivatives of the one-forms \eqref{one-forms_quat} given in \cite{Ferrara:1989ik}, we find
\begin{equation}\label{K_Mink}\begin{aligned}
    \K = \diff \omega^3 = \iu (v \wedge \bar v + u \wedge \bar u + E \wedge \bar E  - e \wedge \bar e)
\end{aligned}\end{equation}
for the K\"ahler two-form. In order to compute the K\"ahler potential, we use \eqref{quat_connection} to determine the holomorphic component of $\omega^3$ to be $\omega^3_a=\tfrac{\iu}{2} (v_a - \partial_a K^{\rm h}) $. Inserting this into \eqref{Kahlerconnection} and integrating finally yields
\begin{equation}\label{special_Kpot}
    \K = K^{\rm h}(\hat z, \bar {\hat z}) + 2 \phi \ .
\end{equation}
The K\"ahler potential $\K$ given in \eqref{special_Kpot} is still expressed in terms of the original $\cN=2$ field variables. We can find the corresponding holomorphic coordinates by starting from the superconformal theory, modding out $k_1$ and $k_2$ as a K\"ahler quotient and projecting at the same time $SU(2)\to U(1)$ in the fibre, as explained at the end of Section \ref{section:NoneK}. This gives a rigid K\"ahler space that is a $U(1)\times \mathbb{R}_+$ fibration over ${\Mq}_{\rm h}$.
Inspired by the holomorphic coordinates on the hyper-K\"ahler cone (and the corresponding twistor space) we make the following ansatz \cite{deWit:1997vg,deWit:1998zg,Rocek:2005ij}:
\begin{equation} \label{holcoords_Min}\begin{aligned}
 w^0  ~= &~\e^{-2\phi} + \iu (\tilde \phi + \xi^A (\tilde \xi_A - {\cal G}_{AB} \xi^B)) \ , \\
 w_A  ~= &~ -\iu (\tilde \xi_A - {\cal G}_{AB} \xi^B) \ ,
\end{aligned}\end{equation}
together with the manifestly holomorphic base coordinates $\hat z^a$. As discussed in Appendix~\ref{section:holcoords}, the coordinates $(z^a, w^0, w_A)$ form complex coordinates with respect to the integrable complex structure $J^3$ on $\M_{\rm h}$.
The third condition in \eqref{constant_Min} then reads
\begin{equation} \label{constant_Min_2}
D^A w_A = \textrm{const.} \ ,
\end{equation}
which is a holomorphic equation on $\M_{\rm h}$.
On the quotient ${\Mq}_{\rm h}$ the coordinates \eqref{holcoords_Min} form equivalence classes under shifts by $k_1$ and $k_2$, i.e.\ under
\begin{equation}\label{equivalence_Min} \begin{aligned}
  w^0 ~\sim & ~ w^0 - 2  \lambda \bar D^A w_A + 2  \lambda \bar D^A \bar w_A \ , \\
  w_A ~\sim & ~ w_A + \iu \lambda {\cal G}_{AB} \bar D^B - \iu \lambda \bar {\cal G}_{AB} \bar D^B \ ,
\end{aligned}\end{equation}
where $\lambda \in \mathbb{C}$ and in both equivalence relations the first shift is holomorphic in the coordinates and the second one is constant due to \eqref{constant_Min_2}. In Appendix \ref{section:holcoords} we show that the coordinates $(w^0,w_A, \hat z^a)$ together with the constraint \eqref{constant_Min} and the identification \eqref{equivalence_Min}
give a set of holomorphic coordinates with respect to $\hat J$. Now we can express $\phi$ and the K\"ahler potential $\K$ in \eqref{special_Kpot} in terms of these holomorphic coordinates via \cite{Rocek:2005ij}
\begin{equation}
 \phi = - \tfrac12 \ln \big( ( w^0 +\bar w^0) + (w_A + \bar w_A) (\Im{\cal G})^{-1\, AB} (w_B + \bar w_B) \big) \ .
\end{equation}

So far we just considered the Killing vectors $k_{1,2}$ given in \eqref{Killingv}. Now let us assume that there are additional gaugings for the remaining Killing vectors $k_{\tilde \lambda} = (k_A, \tilde k^A)$ and $k_{\tilde \phi}$, at a scale well below $\mino$.
The superpotential generated by these can be found by inserting the Killing prepotentials \eqref{prepotential_no_compensator} into the general expression \eqref{Wone}, from which we find
\begin{equation} \label{superpot_Min}
   {\cal W} = 2 V^\Lambda \Theta^{\phantom{\Lambda}\tilde\lambda}_\Lambda U_{\tilde\lambda}  \ ,
\end{equation}
where $U_{\tilde\lambda} = (Z^A , {\cal G}_A)$.
We see that ${\cal W}$ is manifestly holomorphic, consistent with our proof in Section \ref{section:NoneW}.
Furthermore, it depends on the scalars from both the vector- and hypermultiplet sectors.
The D-terms are obtained by insertion of $P^3$ into \eqref{D_terms}. They read
\begin{equation}\label{D_terms_Min}
  {\cal D}^{\hat I} = - \e^{2\phi} (\mathrm{Re} f)^{-1 ~\hat I \hat J} \left( \left( \Theta_{\hat J}^{~~\bar \lambda} - \iu \bar{f}_{\hat J \hat K} \Theta^{\hat K\bar \lambda}\right) \xi_{\bar \lambda}
  - \left( \Theta_{\hat J}^{~~\tilde \phi} - \iu \bar{f}_{\hat J \hat K} \Theta^{\hat K \tilde \phi}\right) \right) \ .
\end{equation}

\subsection{AdS vacua} \label{section:AdS_special}
Let us now determine the K\"ahler potential and the superpotential of effective $\cN=1$ theories that have AdS ground states. In this case the $\cN=2$ supersymmetry parameter that preserves the $\cN=1$
has the form \cite{Louis:2009xd,Grana:2006kf}
\begin{equation}  \label{unbroken_SUSY_AdS}
\epsilon^{\cal A}_1 = (\e^{\iu \varphi/2}, \e^{-\iu \varphi/2})\, \epsilon \ ,
\end{equation}
where $\epsilon$ is the $\cN=1$ generator and $\varphi$ is an arbitrary phase.
In order to use the expressions of the previous section, we first perform an $SU(2)$-rotation given by
\begin{equation}
\epsilon^{\cal A}\to {M}^{\cal A}_{\phantom{\cal A} \cal B}\epsilon^{\cal B}\ ,\qquad \textrm{where}\qquad
{M}^{\cal A}_{\phantom{\cal A} \cal B}= \tfrac{1}{\sqrt{2}}\left( \begin{aligned}
           \e^{\iu\varphi/2} && - \e^{\iu\varphi/2} \\ \e^{-\iu\varphi/2} &&\e^{-\iu\varphi/2}
           \end{aligned}\right) \ .
\end{equation}
This in turn rotates the Killing prepotentials according to
\begin{equation} \label{twisted_prepotentials}
 P^-_{1,2} \to \tilde P^-_{1,2} = \iu \Im(\e^{\iu \varphi} P^-_{1,2}) - P^3_{1,2} \ ,
\end{equation}
and the connection to
\begin{equation} \label{vanishingP3}
\omega^3  \to \tilde \omega^3 = \Re(\e^{\iu \varphi} \omega^-) = 2 \Im(\e^{\iu \varphi} u) \ .
\end{equation}

We have shown in \cite{Louis:2009xd} that the conditions for partial supersymmetry breaking in an AdS ground state are solved by the two Killing vectors
\begin{equation}\label{kv_AdS}
 k_1 = \Re \Big( \e^{\iu\varphi} (Z^A k_A + {\cal G}_A \tilde k^A)\Big) \ , \qquad k_2= k_{\tilde \phi} \ .
\end{equation}
The prefactors of $(k_A,\tilde k^A)$ should again be constant in the $\cN=1$ locus and therefore all coordinates $z^a$ of the base space $\M_{\rm sk}$ are stabilised. It is straightforward to check that the SU(2) gauge choice $\tilde P^3_{1,2}=0$ and $\diff \tilde P^3_{1,2}=0$ holds. We can now use \eqref{Kdef} and \eqref{vanishingP3} to compute the K\"ahler two-form $\K$ on ${\Mq}_{\rm h}$, finding
\begin{equation} \label{K_2form_AdS}\begin{aligned}
  \K =& \ 2\Im\big(\e^{\iu \varphi} u \big) \wedge \Re v - 2 \Im\Big(\e^{\iu \varphi}\bar E \wedge e \Big) \\
  &+ 2 \iu \e^{K^{\rm h}} \Re \big(\e^{\iu \varphi} u\big) \wedge \Big(Z^A (\Im  \mathcal G_{AB})\diff \bar Z^B - \bar Z^A (\Im  \mathcal G_{AB}) \diff Z^B \Big)  \ .
\end{aligned}\end{equation}
With the help of the associated complex structure we then identify the holomorphic part of $\tilde \omega^3$ to be $\tilde \omega^3_a= 2 (\Im(\e^{\iu \varphi} u_a) - \iu (v+\bar v)_a)$. Inserting this into \eqref{Kahlerconnection} leads to the K\"ahler potential
\begin{equation}\label{special_Kpot_AdS}
 \K = 4 \phi \ .
\end{equation}

Analogous to the Minkowski case, one can find holomorphic coordinates on ${\Mq}_{\rm h}$ by going to the corresponding superconformal theory. We use the ansatz \cite{Neitzke:2007ke}
\begin{equation} \label{holcoords_AdS}
 w_{\tilde \lambda} ~=~ \xi_{\tilde \lambda} + 2 \iu \Im \Big(\e^{K^{\rm h}/2-\phi+\iu \varphi} U_{\tilde\lambda} \Big) \ ,
\end{equation}
where $\xi_{\tilde \lambda}=(\xi^A,\tilde \xi_A)$ and $U_{\tilde\lambda}=(Z^A,{\cal G}_A)$.
We will see below that this leads to holomorphic coordinates with respect to $\hat J$ if one imposes the equivalence relation
\begin{equation}
 \xi_{\tilde \lambda} ~\sim~ \xi_{\tilde \lambda} + \lambda \Re \Big( \e^{\iu\varphi} U_{\tilde\lambda} \Big) \ ,
\end{equation}
for any real number $\lambda$.
In terms of $w_{\tilde \lambda}$, the K\"ahler potential \eqref{special_Kpot_AdS} is expressed as
\begin{equation}\label{K_AdS}
  \K = -2 \ln \left( \tfrac14 \Im w_{\tilde \lambda}  G^{\tilde \lambda \tilde \rho} \Im w_{\tilde \rho} \right) \ ,
\end{equation}
where $G^{\tilde \lambda \tilde \rho}$ is the well-known matrix \cite{Ceresole:1995ca}
\begin{equation}
 G^{\tilde \lambda \tilde \rho} = \left( \begin{array}{ccc} (\Im {\cal G})_{AB} + (\Re {\cal G})_{AC}(\Im {\cal G})^{-1\,CD}(\Re {\cal G})_{DB} && -(\Re {\cal G})_{AC} (\Im {\cal G})^{-1\, CB} \\ - (\Im {\cal G})^{-1\, AC} (\Re {\cal G})_{CB} && (\Im {\cal G})^{-1\, AB}  \end{array} \right) \ .
\end{equation}

Inserting the Killing prepotentials \eqref{twisted_prepotentials} and \eqref{prepotential_no_compensator} into the general expression for the superpotential \eqref{Wone}, we obtain
\begin{equation}\label{superpot_AdS}
 {\cal W} = {\cal W}_0 + V^\Lambda \Theta^{\tilde \lambda}_\Lambda w_{\tilde \lambda}  \ ,
\end{equation}
where ${\cal W}_0$ is constant and related to the cosmological constant via $\mu = \e^{K^{\cN=1}/2} {\cal W}_0 $, with $K^{\cN=1}$ evaluated at the $\cN=1$ point. In Section \ref{section:NoneW} we already showed that the superpotential ${\cal W}$ is holomorphic with respect to $\hat J$. Since all coordinates $w_{\tilde \lambda}$ of $\Mq_{\rm h}$ appear in \eqref{superpot_AdS} as coefficients of $\Theta^{\tilde \lambda}_\Lambda$, we can conclude that these coordinates are indeed holomorphic with respect to $\hat J$.

Before we continue, let us consider the case with just $k_{1,2}$ gauged such that ${\cal W} = {\cal W}_0$. Then the $\cN=1$ F-term condition implies that $\K$ is extremal at the supersymmetric minimum and all scalars appearing in $\K$ given in \eqref{K_AdS}, i.e.\ all $\Im w_{\tilde \lambda}$, are stabilised. From \eqref{holcoords_AdS} we then see that the dilaton and all base coordinates are stabilised, consistent with the discussion below \eqref{kv_AdS}.

The D-terms are obtained by insertion of \eqref{prepotential_no_compensator} and \eqref{vanishingP3} into \eqref{D_terms}, resulting in
\begin{equation}\label{D_terms_AdS}\begin{aligned}
  {\cal D}^{\hat I} =& - \e^{2 \phi} (\mathrm{Re} f)^{-1 ~\hat I \hat J}  \left( \Theta_{\hat J}^{~~\tilde \lambda} - \iu \bar{f}_{\hat J \hat K} \Theta^{\hat K\tilde \lambda}\right) \Re(\e^{\iu \varphi} U_{\tilde \lambda}) \\ =& - \e^{2 \phi} (\mathrm{Re} f)^{-1 ~\hat I \hat J}  \left( \Theta_{\hat J}^{~~\tilde \lambda} - \iu \bar{f}_{\hat J \hat K} \Theta^{\hat K\tilde \lambda}\right) G_{\tilde \lambda}^{~~ \tilde \rho} \Im(w_{\tilde \rho})\ ,
\end{aligned}\end{equation}
where we lowered one index in $G_{\tilde \lambda}^{~~ \tilde \rho}$ by using the standard symplectic metric.

Finally, let us note that the K\"ahler potential $\K$ \eqref{special_Kpot_AdS} also coincides with the expression obtained in orientifold truncations of the type II compactifications considered in \cite{Benmachiche:2006df} (see also \cite{Cassani:2007pq} and references therein). This is expected from the form of $S_{\cal AB}$ in supergravities with special quaternionic-K\"ahler $\M_{\rm h}$ when the unbroken $\cN=1$ supersymmetry generator has the form \eqref{unbroken_SUSY_AdS} \cite{Grana:2005ny,Grana:2006hr}. Furthermore, \eqref{superpot_AdS} is similar to the superpotential derived in \cite{Shelton:2006fd,Aldazabal:2006up,Micu:2007rd,Cassani:2007pq} for $\cN=1$ truncations of $\cN=2$ supergravity, up to the directions we have integrated out.

\section{Conclusions}\label{Conc}

We have derived the $\cN=1$ low-energy effective action of partially
broken $\cN=2$ gauged supergravity. We first kept the analysis as general as
possible, in that we only assumed the existence of maximally-symmetric $\cN=1$ backgrounds
without further specifying any particular supergravity. This implies
that the $\cN=2$ spectrum has to contain electrically and magnetically
charged hypermultiplets arising from two commuting isometries on
the quaternionic-K\"ahler manifold $\M_{\rm h}$.
The corresponding
Killing vectors can be combined into one complex Killing vector which
has to be holomorphic with respect to one of the three almost complex
structures of $\M_{\rm h}$. For
this class of supergravities we explicitly computed the couplings of
the $\cN=1$ low-energy effective action in terms of the original
$\cN=2$ `data' and showed their consistency with the general
constraints of $\cN=1$ supersymmetry.

The main issue in checking the $\cN=1$ supersymmetry of the low-energy effective
theory is related to the necessary K\"ahler
property of the scalar field space. Although the component $\M_{\rm
  h}$ of the original $\cN=2$ field space is not a K\"ahler manifold,
the quotient $\Mq_{\rm h}$ arising from integrating out the two heavy
gauge bosons is K\"ahler. The dimension of this quotient depends on
the details of the theory, but can be as large as $(4n_{\rm h}-2)$,
where only the two Goldstone bosons have been removed. However, generically a large number of moduli are fixed leaving a low-dimensional $\cN=1$ field space. This differs from truncated theories where the scalar field space is a submanifold of maximal dimension $2n_{\rm h}$, in agreement with the mathematical results of \cite{Alekseevsky}. Thus, our quotient construction is an interesting mathematical result in itself, which we shall further expand on in a companion paper \cite{Cortes}.

Once the K\"ahler structure is identified it is relatively straightforward to also check the holomorphicity of the superpotential, which we confirmed in Section~\ref{section:NoneW}. We found that the holomorphicity of the gauge couplings was a consequence of integrating out the graviphoton, which is necessarily one of the heavy gauge bosons. Similarly, in Section \ref{section:NoneD} we saw that the restriction to the light gauge bosons led to the correct form of the $\cN=1$ D-term. Finally, in Section \ref{section:SQC} we gave an example of our construction by deriving the $\cN=1$ K\"ahler potential, the superpotential and the D-terms arising from partial supersymmetry breaking in an $\cN=2$ supergravity with a special quaternionic-K\"ahler manifold for both Minkowski and AdS backgrounds. For this example we argued that a large number of moduli are stabilised

\vskip 1cm

\subsection*{Acknowledgements}

This work was supported by the German Science Foundation (DFG) under the Collaborative Research Center (SFB) 676. We have greatly benefited from conversations and correspondence with P.\ Berglund, D.\ Cassani, S.\ Cecotti, V.\ Cort\'es, G.\ Dall'Agata, C.\ Scrucca, S.~Vandoren and E.\ Zaslow.

\vskip 1cm

\appendix
\noindent
{\bf\Large Appendix}

\section{The massive spin-3/2 multiplet} \label{section:massive_multi}

Here we will compute the normalised masses of the gravitino and the two gauge bosons in the massive spin-3/2 multiplet and show their consistency with the $\cN=1$ mass relations. It is well known that when supersymmetry is spontaneously broken, one massless combination of the spin-$1/2$ fields, the Goldstino, gets eaten by
the gravitino. For the case of spontaneous partial supersymmetry breaking, we can identify the Goldstino by its coupling to the gravitino $\Psi_{\mu 2}$ corresponding to the broken supersymmetry generator $\epsilon_2$ in the fermionic Lagrangian, which should be of the form $\bar\eta\gamma^\mu\Psi_{\mu 2}~$ \cite{Volkov:1973ix,Deser:1977uq,Cremmer:1982wb}. By
examining the fermionic contributions to the $\cN=2$ supergravity Lagrangian \cite{Andrianopoli:1996cm}, one can see that the Goldstino should be
defined by \cite{Gunara:2003td}
\begin{equation}
\eta =  g_{\bar i j} \bar{W}^{\bar{i}}_{2 \cal{A}} \lambda^{j\cal{A}} + 2 \bar N_2^\alpha\zeta_\alpha~.
\end{equation}
The Goldstino is then removed from the Lagrangian by an appropriately
chosen gauge transformation of $\Psi_{\mu 2}$.
This is the superHiggs effect.
The mass term of $\Psi_{\mu 2}$ can be read off from the components of
$S_{\cal AB}$ appearing in its supersymmetry variation
\eqref{susytrans2}.
For our choice of the preserved supersymmetry parameter $\epsilon_1 =
{\epsilon\choose  0}$ one finds\footnote{Here and in the following, we omit the label ``AdS'' on $C^I$ for the case of the AdS solution \eqref{solution_embedding_tensor_AdS2}.}
\begin{equation} \label{mass_grav}
 m^2_{\Psi_{\mu 2}} = 16 \bar S^{22} S_{22} = 4 \e^{K^{\rm v}} \bar C^I \Im {\cal F}_{IJ} X^J \bar X^K \Im {\cal F}_{KL} C^L (P^1_1 P^2_2 - P^1_2 P^2_1) \ ,
\end{equation}
where we have used \eqref{susytrans3} and the embedding tensor solutions \eqref{solution_embedding_tensor2} or \eqref{solution_embedding_tensor_AdS2}.

The residual $\cN=1$ supersymmetry implies that the massive gravitino
should reside in a complete, massive spin-$3/2$ multiplet with spin
content $(3/2, 1, 1, 1/2)$. In Section~\ref{section:NoneK} we identified the
two massive gauge bosons \eqref{Acomb} which should lie in this
multiplet. As an additional consistency check, we shall now explicitly
compute their masses and confirm that they agree with the gravitino mass
\eqref{mass_grav}, as required by $\cN=1$ supersymmetry.

In order to compute the masses for the two heavy vectors it is
convenient to go to a purely electric frame. This is facilitated by an $Sp(n_{\mathrm v}+1)$ transformation
\begin{equation}
 \tilde\Theta^\Lambda = {\cal S}^\Lambda_{~\Sigma} \Theta^\Sigma\ ,
\end{equation}
where $\Theta^\Sigma$ is given in \eqref{solution_embedding_tensor_AdS2} and
\begin{equation}
  {\cal S}^\Lambda_{~\Sigma} \ = \left(
    \begin{array}{cc}
      U^I_{~J} & Z^{IJ} \\
      [1mm] W_{IJ} & V^{~J}_I
    \end{array}
  \right) \ ,
\end{equation}
whose submatrices obey
\begin{equation}\begin{aligned}
  \label{spc2}
  U^{\rm T} V- W^{\rm T} Z &= V^{\rm T}U - Z^{\rm T}W =
  {\bf 1}\, ,\\
  U^{\rm T}W = W^{\rm T}U\,, & \quad Z^{\rm T}V= V^{\rm T}Z\ .
\end{aligned}\end{equation}
Demanding both rotated charges $\tilde\Theta_1$ and $\tilde\Theta_2$ to be purely electric implies the following conditions:
\begin{equation}\label{charges_rot_magn}
 \tilde\Theta_1^I + \iu \tilde\Theta_2^I =(U^I_{~J} + Z^{IK} {\cal F}_{KJ}) C^J + \iu  \e^{K^{\mathrm v}/2} \tfrac{\mu}{P^-_1} (U^I_{~J} + Z^{IK} \bar {\cal F}_{KJ}) \bar X^J = 0 \ .
\end{equation}
The electric charges in the rotated frame are given explicitly by
\begin{equation}\label{charges_el_rot}\begin{aligned}
 \tilde \Theta_{1 \, I} ~= &~ \Re \big( (W_{IJ}+V^{~K}_I {\cal F}_{KJ}) (C^J + \iu  \e^{K^{\mathrm v}/2} \tfrac{\bar \mu}{P^+_1} X^J) \big) \ , \\
 \tilde \Theta_{2 \, I} ~= &~ \Im \big( (W_{IJ}+V^{~K}_I {\cal F}_{KJ}) (C^J - \iu  \e^{K^{\mathrm v}/2} \tfrac{\bar \mu}{P^+_1} X^J) \big) \ .
\end{aligned}\end{equation}
Note that one recovers the charges \eqref{solution_embedding_tensor_AdS2} in the original frame by applying the inverse transformation, i.e.\
\begin{equation}\label{charges_rot_back}\begin{aligned}
 \Theta_{1 \, I} ~= &~ \quad (U^{\rm T})^{~J}_I \tilde \Theta_{1 \, J} =  \Re \big({\cal F}_{IJ} ( C^J - \iu \e^{K^{\rm v}/2} \tfrac{\bar \mu}{P^+_1} X^J) \big) \ , \\
 \Theta_{1}^{\, I} ~= &~ -(Z^{\rm T})^{IJ} \tilde \Theta_{1 \, J} =  \Re \big( C^I - \iu \e^{K^{\rm v}/2} \tfrac{\bar \mu}{P^+_1} X^I \big) \ , \\
\Theta_{2 \, I} ~= &~ \quad (U^{\rm T})^{~J}_I \tilde \Theta_{2 \, J} =  \Im \big({\cal F}_{IJ} ( C^J + \iu \e^{K^{\rm v}/2} \tfrac{\bar \mu}{P^+_1} X^J) \big) \ , \\
\Theta_{2}^{\, I} ~= &~ -(Z^{\rm T})^{IJ} \tilde \Theta_{2 \, J} =  \Im \big( C^I + \iu \e^{K^{\rm v}/2} \tfrac{\bar \mu}{P^+_1} X^I \big) \ .
\end{aligned}\end{equation}

The masses of the two heavy gauge bosons can then be read off from the scalar covariant derivative terms in the
Lagrangian \eqref{sigmaint} with \eqref{d2} inserted. In the purely electric frame they take the form
\begin{equation}\label{vector_mass_term}
 {\cal L}_\textrm{mass} = \Big(h(k_1,k_1)\tilde \Theta_{1 \, I} \tilde \Theta_{1 \, J} + h(k_2,k_2)\tilde \Theta_{2\, I} \tilde \Theta_{2 \, J}\Big)  A^I_\mu A^{J\mu} \ ,
\end{equation}
where $h(k_\lambda,k_\rho)\equiv h_{uv}k^u_\lambda k^v_\rho$ and we have already
used $h(k_1,k_2)=0$, which follows from \eqref{Jk}.
In order to compare these expressions with the gravitino mass
\eqref{mass_grav} we have to canonically normalise the gauge
boson kinetic terms.
The gauge kinetic function in the rotated electric frame is given by
$\Im (\tilde {\cal N})_{IJ}$, where \cite{Andrianopoli:1996cm}
\begin{equation} \label{N_rot}
\tilde {\cal N}_{IJ} = (W+V {\cal N}) (U + Z {\cal N})^{-1} \ .
\end{equation}
We can identify the kinetic terms of the massive gauge vectors by projecting ${\cal L}_\textrm{kin}$ onto the subspace spanned by $\tilde \Theta_{1 \, I}$ and $\tilde \Theta_{2 \, I}$, which yields
\begin{equation}\label{vector_kin_term}
 {\cal L}_\textrm{kin} \supset \Big( \frac{\tilde \Theta_{1 \, I} \tilde \Theta_{1 \, J}}{\tilde \Theta_{1 \, K} (\Im \tilde {\cal N})^{-1\,KL}\tilde \Theta_{1 \, L}} + \frac{\tilde \Theta_{2\, I} \tilde \Theta_{2 \, J}}{\tilde \Theta_{2 \, K} (\Im \tilde {\cal N})^{-1\,KL}\tilde \Theta_{2 \, L}}\Big)\,  F^I_{\mu\nu} F^{J\,\mu\nu} \ .
\end{equation}
By comparing the mass \eqref{vector_mass_term} with the kinetic terms
\eqref{vector_kin_term}, we can read off the canonically normalised mass parameters
of the heavy vectors to be
\begin{equation}\label{mass_vector_1}
 m_1^2 = 2 h(k_1,k_1) \tilde \Theta_{1 \, I} \Im (\tilde {\cal N})^{-1\,IJ} \tilde \Theta_{1 \, J} \ , \qquad m_2^2 = 2 h(k_2,k_2) \tilde \Theta_{2 \, I} \Im (\tilde N)^{-1\,IJ} \tilde \Theta_{2 \, J} \ .
\end{equation}
Note that from \eqref{massrel} and \eqref{mass_prepot} we also know that
\begin{equation}\label{quadrelation_k_P}
 h(k_1,k_1)= h(k_2,k_2) = \tfrac12 (P^1_1 P^2_2 - P^1_2 P^2_1) \ .
\end{equation}

In order to compare the vector masses with the gravitino mass
 \eqref{mass_grav} we need to explicitly compute
$\tilde \Theta_{1,2 \, I} \Im (\tilde {\cal N})^{-1\,IJ} \tilde \Theta_{1,2 \, J}$. To do this, we will use the decomposition
$  C^I = C^{(Z)\,I} + C^{(P)\,I} $ where
\begin{equation}
C^{(Z)\,I} = -2\e^{K^{\rm v}} C^J \Im {\cal F}_{JK} \bar X^K X^I  \ ,\qquad
 C^{(P)\,I} = C^J {\Pi}_J^{\phantom{J}I} \ ,
\end{equation}
and ${\Pi}_J^{\phantom{J}I}$ was defined in \eqref{Pidef}. By definition, the following relations hold
\begin{equation} \label{decompose_properties}
 C^{(Z)\,I} {\cal F}_{IJ} = C^{(Z)\,I} {\cal N}_{IJ} \ , \qquad C^{(P)\,I} {\cal F}_{IJ} = C^{(P)\,I} \bar{\cal N}_{IJ} \ .
\end{equation}
Using these relations the electric charges in \eqref{charges_el_rot} can be written as
\begin{equation}\label{charges_el_rot_rewrite}\begin{aligned}
 \tilde \Theta_{1 \, I}~ =~ & \Re \Big(\tilde {\cal N}_{IL} (U^L_{~J} + Z^{LK} {\cal F}_{KJ}) (C^{(Z)\,J} - \iu  \e^{K^{\mathrm v}/2} \tfrac{\bar \mu}{P^+_1} X^J) \\ & \quad \ + {\bar{\tilde {\cal N}}}_{IL} (U^L_{~J} + Z^{LK} {\cal F}_{KJ}) C^{(P)\,J} \Big) \ , \\
 \tilde \Theta_{2 \, I}~ =~ & \Im \Big( \tilde {\cal N}_{IL} (U^L_{~J} + Z^{LK} {\cal F}_{KJ}) (C^{(Z)\,J} + \iu  \e^{K^{\mathrm v}/2} \tfrac{\bar \mu}{P^+_1} X^J) \\ & \quad \ + {\bar{\tilde {\cal N}}}_{IL} (U^L_{~J} + Z^{LK} {\cal F}_{KJ}) C^{(P)\,J} \Big) \ ,
\end{aligned}\end{equation}
where we have factored out $\tilde {\cal N}_{IL}$ given in \eqref{N_rot}. If we now use \eqref{charges_rot_magn} and \eqref{decompose_properties}, we arrive at
\begin{equation}\label{charges_el_rot_rewrite2}\begin{aligned}
 \tilde \Theta_{1 \, I}~= &~ -2 (\Im \tilde {\cal N})_{IL} \Im \big( (U^L_{~J} + Z^{LK} {\cal F}_{KJ}) C^{(Z)\,J}  \big) \ , \\
 \tilde \Theta_{2 \, I}~= &~ \quad 2 (\Im \tilde {\cal N})_{IL} \Re \big( (U^L_{~J} + Z^{LK} {\cal F}_{KJ}) C^{(Z)\,J} \big) \ .
\end{aligned}\end{equation}
Combining \eqref{charges_el_rot_rewrite2} with
\eqref{charges_rot_back}, we find
\begin{equation}\label{Theta_12_squared}\begin{aligned}
 \tilde \Theta_{1 \, I} \Im (\tilde {\cal N})^{-1\,IJ} \tilde \Theta_{1 \, J}~ =~ & -2 \Im \Big( C^{(Z)\,I} \big( \Re \big({\cal F}_{IJ} ( C^J + \iu \e^{K^{\rm v}/2} \tfrac{\bar \mu}{P^+_1} X^J)\big) \\ & \phantom{-2 \Im \Big( C^{(Z)\,I} \big( }
 - {\cal N}_{IJ} \Re \big( C^J + \iu \e^{K^{\rm v}/2} \tfrac{\bar
\mu}{P^+_1} X^J \big) \big) \Big) \\ ~ =~ & 2 C^{(Z)\,I} \Im {\cal N}_{IJ} \bar C^{(Z)\,J} + 2 \Im \big( \e^{K^{\rm v}/2} \tfrac{\mu}{P^-_1} C^{(Z)\,I} \Im {\cal N}_{IJ} \bar X^J \big) \ , \\
  \tilde \Theta_{2 \, I} \Im (\tilde {\cal N})^{-1\,IJ} \tilde
\Theta_{2 \, J} ~ =~ & 2 \Re \Big( C^{(Z)\,I} \big( \Im \big({\cal F}_{IJ} ( C^J + \iu \e^{K^{\rm v}/2} \tfrac{\bar \mu}{P^+_1} X^J)\big) \\ & \phantom{2 \Re \Big( C^{(Z)\,I} \big( }
  - {\cal N}_{IJ} \Im \big( C^J + \iu \e^{K^{\rm v}/2} \tfrac{\bar
\mu}{P^+_1} X^J \big) \big) \Big) \\ ~ =~ & 2 C^{(Z)\,I} \Im {\cal N}_{IJ} \bar C^{(Z)\,J} - 2 \Im \big( \e^{K^{\rm v}/2} \tfrac{\mu}{P^-_1} C^{(Z)\,I} \Im {\cal N}_{IJ} \bar X^J \big) \ .
\end{aligned}\end{equation}
By using \eqref{Theta_12_squared} and \eqref{quadrelation_k_P} we can determine the vector boson masses \eqref{mass_vector_1} to be
\begin{equation}\label{mass_vector_2}
 m_{1,2}^2 = 2\left( C^{(Z)\,I} \Im {\cal N}_{IJ} \bar C^{(Z)\,J} \pm \Im \big( \e^{K^{\rm v}/2} \tfrac{\mu}{P^-_1} C^{(Z)\,I} \Im {\cal N}_{IJ} \bar X^J \big) \right) (P^1_1 P^2_2 - P^1_2 P^2_1) \ .
\end{equation}

In a Minkowski background we have $\mu=0$ and we see that the vector masses are equal to the gravitino mass \eqref{mass_grav}, which can be rewritten as
\begin{equation}
 m^2_{\Psi_{\mu 2}} = -\tfrac12  \bar C^{(Z)\,I} \Im {\cal N}_{IJ} C^{(Z)\,J} (P^1_1 P^2_2 - P^1_2 P^2_1) \ .
\end{equation}

For an AdS background $\mu \ne 0$ and so the masses differ. Let us define $l^2= |\mu|^2$. From the representation theory of AdS superalgebras one finds that a massive spin-$3/2$
multiplet obeys the following mass relations \cite{Castellani:1991et}
\begin{equation}\label{AdS_mass_pattern}\begin{aligned}
(m_{\psi_\mu})^2~ =~ & (m_{3/2}-l)^2~ =~  m^2~,\\
m_1^2~ =~ & m(m-l)~, \\
m_2^2~ =~ & m(m+l)~, \\
m_{1/2}~ =~ & m^2~.
\end{aligned}\end{equation}
This agrees with the mass splitting we have found in \eqref{mass_vector_2}, if we identify
\begin{equation}
 m^2=\e^{K^{\rm v}} C^{(Z)\,I} \Im {\cal N}_{IJ} \bar C^{(Z)\,J}(P^1_1 P^2_2 - P^1_2 P^2_1) \ .
\end{equation}

\section{Holomorphic coordinates}
\label{section:holcoords}

In this appendix we prove the holomorphicity of the coordinates in the $\cN=1$ low-energy effective theory for a Minkowski background. To do so, we first show the holomorphicity of the coordinates $(z^a,w^0,w_A)$ introduced in \eqref{holcoords_Min}
with respect to $J^3$ on $\M_{\mathrm{h}}$. In fact this shows that $J^3$ is 
already integrable on $\M_{\mathrm{h}}$.\footnote{For the Wolf spaces this was shown in \cite{Cortesthesis} but the proof can be generalised to all quaternionic-K\"ahler spaces which are in the image of the c-map. We thank V.~Cort\'es for discussions on this issue.} From the construction of $\Mq_{\mathrm{h}}$ we then see that the coordinates descend to holomorphic coordinates with respect to $\hat J$ on $\Mq_{\mathrm{h}}$.
The base coordinates $z^a$ on $\M_{\mathrm{h}}$ are manifestly holomorphic with respect to $J^3$ and thus they straightforwardly 
descend to  holomorphic coordinates 
with respect to $\hat J$ on $\Mq_{\mathrm{h}}$. Therefore, we focus on $w^0$ and $w_A$, given in \eqref{holcoords_Min}, and show that these coordinates are also holomorphic coordinates with respect to $J^3$ on $\M_{\mathrm{h}}$. This is done by computing their exterior derivatives and showing that they give $(1,0)$-forms on $\M_{\mathrm{h}}$.

The exterior derivative of $w_A$ is
\begin{equation}
 \diff w_A = -\iu (\diff \tilde \xi_A - {\cal G}_{AB} \diff \xi^B) - \iu {\cal G}_{ABC} \xi^B \diff Z^C  \ .
\end{equation}
The last term is clearly a $(1,0)$-form as $Z^A$ is a holomorphic function of the $z^a$. Using the definition of the one-forms \eqref{one-forms_quat} and the identity \cite{Ferrara:1989ik}
\begin{equation}\label{sumtounit}
\delta_A^B = ~\tfrac12 \e^{-K^{\rm h}} \Pi_{A \underline{b}} \bar \Pi_C^{~\underline{b}} (\Im {\cal G})^{-1\,CB} - 2 \e^{K^{\rm h}} (\Im {\cal G})_{AC} \bar Z^C Z^B \ ,
\end{equation}
we find
\begin{equation}\label{hol_fibre_one-forms}
\diff \tilde \xi_A - {\cal G}_{AB} \diff \xi^B = -\iu \e^{-K^{\rm h}/2 - \phi} \Pi_{A \underline{b}} \bar E^{\underline{b}} + 2 \iu \e^{K^{\rm h}/2 - \phi} (\Im {\cal G})_{AC} \bar Z^C u \ ,
\end{equation}
where we used the standard relations
\begin{equation}
 \bar \Pi_A^{~\underline{b}} (\Im {\cal G})^{-1\,AB} {\cal G}_{BC} =  \bar \Pi_A^{~\underline{b}} (\Im {\cal G})^{-1\,AB} \bar{\cal M}_{BC} \ , \qquad Z^A {\cal G}_{AB} = Z^A {\cal M}_{AB} \ ,
\end{equation}
which follow from the definition of $\mathcal M_{AB}$ (analogous to \eqref{Ndef}). From the definition of the vielbein \eqref{quat_vielbein} we know that $(u,\bar v, e,\bar E)$ are $(1,0)$-forms of $J^3$, therefore we conclude from \eqref{hol_fibre_one-forms} that $\diff \tilde \xi_A - {\cal G}_{AB} \diff \xi^B$ and thus $\diff w_A$ are indeed $(1,0)$-forms.

It remains to show that the exterior derivative of $w^0$ is a $(1,0)$-form with respect to $J^3$. From \eqref{holcoords_Min} we find
\begin{equation}\begin{aligned}
\diff w^0 = &~ \diff \e^{-2\phi} + \iu \diff \tilde \phi + \iu \diff\xi^A (\tilde \xi_A - {\cal G}_{AB} \xi^B) + \iu \xi^A (\diff \tilde \xi_A - {\cal G}_{AB} \diff \xi^B) + \iu \xi^A {\cal G}_{ABC} \xi^B \diff Z^C
\\ = &~ 2 \e^{-2\phi} \bar v + 2 \iu \xi^A (\diff \tilde \xi_A - {\cal G}_{AB} \diff \xi^B) + \iu \xi^A {\cal G}_{ABC} \xi^B \diff Z^C \ ,
\end{aligned}\end{equation}
where we have used \eqref{one-forms_quat}. Again, $\bar v$ and the last term are clearly $(1,0)$-forms and we have already shown that $\diff \tilde \xi_A - {\cal G}_{AB} \diff \xi^B$ is a $(1,0)$-form. Thus, $\diff w^0$ is a $(1,0)$-form with respect to $J^3$ and $w^0$ is a 
holomorphic coordinate.

To summarise, in this appendix 
we have shown that $(z^a, w^0, w_A)$ locally define a set of holomorphic coordinates with respect to $J^3$ on $\M_{\rm h}$. One can furthermore check that the fibre coordinates $(w^0, w_A)$ transform holomorphically under chart transitions in the base $\M_{\rm sk}$, due to the transformation properties of $(\xi^A,\tilde \xi_A)$ and ${\cal G}_{AB}$ under symplectic rotations. Therefore, the complex structure $J^3$ is integrable and admits a set of holomorphic coordinates $(z^a, w^0, w_A)$.
As \eqref{equivalence_Min} defines a quotient with respect to a holomorphic coordinate and \eqref{constant_Min_2} gives a holomorphic subspace, we see that on $\Mq_{\rm h}$ a subset of $(z^a, w^0,w_A)$ gives holomorphic coordinates with respect to the complex structure $\hat J$ constructed in Section \ref{section:NoneK}. A similar computation should be possible in the AdS case.

\bibliographystyle{JHEP-2}

\providecommand{\href}[2]{#2}\begingroup\raggedright\endgroup

\end{document}